\documentstyle[11pt]{article}

\if@twoside\oddsidemargin25mm
\else\oddsidemargin25mm\evensidemargin25mm\marginparwidth25mm\fi%
\newcommand{\be}{\begin{equation}}
\newcommand{\ee}{\end{equation}}
\newcommand{\bea}{\begin{eqnarray}}
\newcommand{\eea}{\end{eqnarray}}
\newcommand{\bref}[1]{(\ref{#1})}
\newcommand{\eps}{\epsilon}
\newcommand{\veps}{\varepsilon}

\newcommand{\der}[2]{\frac{\partial #1}{\partial #2}}
\newcommand{\lder}[2]{\frac{\partial_l#1}{\partial #2}}
\newcommand{\rder}[2]{\frac{\partial_r#1}{\partial #2}}
\newcommand{\gh}[1]{{\cal #1}}

\newfont{\nice}{eufm10 scaled\magstep1}

\makeatletter

\def\restric#1#2{{\left. #1 \right|_{#2}}}
\def\dif{{\rm d}}
\def\deriv{\@ifnextchar[{\@deriv}{\@deriv[]}}
   \def\@deriv[#1]#2#3{\mathchoice%
{{\dif^{#1}#2\over\dif{#3}^{#1}}}{{\dif^{#1}#2/\dif{#3}^{#1}}}%
{{\dif^{#1}#2\over\dif{#3}^{#1}}}{{\dif^{#1}#2/\dif{#3}^{#1}}}}
\def\derpar#1#2{\mathchoice%
{{\partial#1\over\partial#2}}{{\partial#1/\partial#2}}%
{{\partial#1\over\partial#2}}{{\partial#1/\partial#2}}}

\def\secteqno{\@addtoreset{equation}{section}%
\def\theequation{\thesection.\arabic{equation}}}

\def\endsecteqno{\def\theequation{\@ifundefined{chapter}%
{\arabic{equation}}{\thechapter.\arabic{equation}}}}

\newcounter{subequation}
\def\thesubequation{\alph{subequation}}
\def\sneqnarray{\stepcounter{equation}\let\@currentlabel=\theequation
\setcounter{subequation}{1}
\def\@eqnnum{{\rm (\theequation.\thesubequation)}}
\global\@eqcnt\z@\tabskip\@centering\let\\=\@eqncr\let\@@eqncr=\@@sneqncr
$$\halign to \displaywidth\bgroup\@eqnsel\hskip\@centering
 $\displaystyle\tabskip\z@{##}$&\global\@eqcnt\@ne
 \hskip 2\arraycolsep \hfil${##}$\hfil
 &\global\@eqcnt\tw@ \hskip 2\arraycolsep $\displaystyle\tabskip\z@{##}$\hfil
  \tabskip\@centering&\llap{##}\tabskip\z@\cr}
\def\endsneqnarray{\@@sneqncr\egroup $$\global\@ignoretrue}
\def\@@sneqncr{\let\@tempa\relax
   \ifcase\@eqcnt \def\@tempa{& & &}\or \def\@tempa{& &}
   \else \def\@tempa{&}\fi
     \@tempa \if@eqnsw\@eqnnum\stepcounter{subequation}\fi
     \global\@eqnswtrue\global\@eqcnt\z@\cr}

\def\nobiblabels{\def\@lbibitem[##1]##2{\@bibitem{##2}}}

\makeatother

\secteqno

\begin{document}

\begin{titlepage}

\begin{flushright}

KUL-TF-93/50\\
UB-ECM-PF 93/14\\
UTTG-16-93\\
January 1994\\

\end{flushright}

\begin{center}

{\LARGE\bf Anomalies and Wess-Zumino terms in an
           extended, regularized Field-Antifield formalism}\\

\vskip 12.mm

{\sc Joaquim Gomis$^{\flat 1}$ and Jordi Par\'{\i}s$^{\sharp 2}$}

\vskip 0.4cm

$^\flat$ \small{\it{Theory Group, Department of Physics}}\\
\small{\it{The University of Texas at Austin}}\\
\small{\it{RLM\,5208, Austin, Texas}}\\
\small{\it{and}}\\
\small{\it{Departament d'Estructura i Constituents de la
           Mat\`eria}}\\
\small{\it{Facultat de F\'{\i}sica, Universitat de Barcelona}}\\
\small{\it{Diagonal 647, E-08028 Barcelona}}\\
\small{\it{Catalonia}}
\\[0.4cm]

$^\sharp$ \small{\it{Instituut voor Theoretische Fysica}}\\
\small{\it{Katholieke Universiteit Leuven}}\\
\small{\it{Celestijnenlaan 200D}}\\
\small{\it{B-3001 Leuven, Belgium}}\\
[1.5cm]

{\bf Abstract}

\end{center}

\begin{quote}

Quantization of anomalous gauge theories with closed, irreducible gauge
algebra within the extended Field-Antifield formalism is further
pursued. Using a Pauli-Villars (PV) regularization of the generating
functional at one loop level, an alternative form for the anomaly is found
which involves only the regulator. The analysis
of this expression allows to conclude that
recently found ghost number one cocycles with nontrivial antifield
dependence can not appear in PV regularization.
Afterwards, the extended Field-Antifield formalism is further completed by
incorporating quantum effects of the extra variables, i.e., by
explicitly taking into account the regularization of the extra sector.
In this context, invariant PV regulators
are constructed from non-invariant ones,
leading to an alternative interpretation of the Wess-Zumino
action as the local counterterm relating invariant and non-invariant
regularizations.
Finally, application of the above ideas to the bosonic string reproduces
the well-known Liouville action and the shift
$(26-D)\rightarrow(25-D)$ at one loop.

\vspace{10mm}

\hrule width 5.cm

{\small
\noindent $^1$
Permanent adress: Dept.\ d'Estructura i
Constituents de la Mat\`{e}ria, U. Barcelona.\\
E-mail: QUIM@EBUBECM1\\
\noindent $^2$
Wetenschappelijk Medewerker, I.I.K.W., Belgium.\\
E-mail: Jordi=Paris\%tf\%fys@cc3.kuleuven.ac.be}

\normalsize
\end{quote}

\end{titlepage}

\section{Introduction}

\hspace{\parindent}%
One of the fundamental aims of the Field-Antifield (FA) formalism
\cite{BV81} is to provide a general framework for the covariant
path integral quantization of gauge theories at Lagrangian level, using as
principal requirement BRST invariance.
In this spirit FA encompasses previous ideas and
developments for quantizing gauge systems based on BRST symmetry
\cite{BRS} \cite{T} \cite{Zinn} and extends them to more complicated
situations (open algebras, reducible systems, etc.).

The requirement of BRST invariance seems to define the full quantum theory
by means of a single equation, the so-called {\it quantum master equation}.
In \cite{TNP89} \cite{T1}, a Pauli-Villars (PV) regularization scheme at
one loop level was introduced to deal with this equation and
anomalies were recognized to arise whenever
it can not be solved in a local way.
Anomalous gauge theories appear thus characterized by the
breakdown of its classical gauge or BRST structure due to quantum
corrections, leading to the fact that some classical pure gauge degrees of
freedom become propagating at quantum level. Therefore,
according to the spirit of covariant quantization of gauge theories,
the convenience arises of developing an extended
formalism which describes in a BRST invariant manner this phenomenon by
the introduction of extra degrees of freedom already at classical level
\cite{FS84}.

A proposal to consistently quantize in a BRST invariant way
anomalous gauge theories with closed irreducible gauge algebras
along FA ideas was considered in \cite{GP93}. There,
by extending the original
configuration space with the addition of extra degrees of freedom,
a solution for the original regularized
quantum master equation at one loop was given in terms of the antifield
independent part of the anomalies.
This solution turned out to be the contribution
of the original and ghost fields to the Wess-Zumino term. However,
questions as regularization of the extra divergent pieces coming from
the new fields as well as their contribution to the Wess-Zumino
action were not considered. Instead,
only a formal BRST invariant measure was constructed for them.

In this paper, we further pursue the program started in \cite{GP93} along
the lines sketched in \cite{Mars}. A brief account of the
FA formalism, using the concepts of classical and gauge-fixed basis
\cite{VP93} \cite{TP93}, is presented in sect.2. In
this way, the quantum master equation naturally appears as a potential
obstruction for the fulfillment of the Slavnov-Taylor identity \cite{ST72}
associated to the BRST symmetry for the effective action, generalizing the
original proposal of Zinn-Justin \cite{Zinn} for Yang-Mills theories.
The analysis is first presented in a formal fashion
and, afterwards, introducing a PV type
regularization at one loop level for the generating functional \cite{T1}.
The regulated BRST Ward identity yields then an alternative expression for
the anomaly involving only the regulator, which is shown equivalent to that
obtained in \cite{TNP89} \cite{T1} in appendix A.
This alternative expression turns out to be
very useful in the analysis performed in sect.3
of the form of the complete anomaly in the space of fields and antifields.
An important result coming out of this analysis is that
recently found ghost number one cocycles with nontrivial antifield
dependence \cite{Brandt} are ruled out by PV regularization.
The extended formalism presented in \cite{GP93}
is further extended in sect.4 by discussing how
regularization for the new sector of variables should proceed for a
specific type of theories. After that, in sect.5, the construction of
(antifield independent) invariant regulators
in the extended configuration space from non-invariant ones is described
and a method to obtain Wess-Zumino actions from
integration of anomalies is proposed. The procedure relies on the form of
the counterterm relating the anomalies
coming from different regulators analyzed in appendix B.
Sect.6 deals with the application of the above ideas to the bosonic string
and sect.7 with the conclusions.
Finally, in appendix C we discuss the transformation properties of the
regulator for the extra variables sector.

\section{Regularized Field-Antifield formalism}

\hspace{\parindent}%
The Field-Antifield formalism is a powerful method for the
study of gauge theories. (for a review, see \cite{rev}).
At the classical level, it can be seen as a general algorithm
to derive a gauge-fixed action $S_\Sigma(\Phi)$ and its BRST
transformation $\delta_\Sigma$ out of a given
classical gauge action $S_0(\phi)$ and its associated gauge structure.
At the quantum level, it provides the tools
to study to what extent this classical BRST symmetry and its underlying
structure are preserved (or not) by quantum corrections. This quantum BRST
structure is further used to study unitarity and renormalizability issues.

\subsection{Classical theory. Classical basis versus Gauge-fixed basis}

\hspace{\parindent}%
Assume $S_0(\phi^i)$ to be a classical
action, invariant under the (infinitesimal) gauge transformations
\be
   \delta\phi^i= R^i_\alpha\veps^\alpha,
   \quad\quad i=1,\ldots,n;\quad\quad \alpha=1,\ldots,m.
\label{orig gau}
\ee
The BV approach starts by enlarging the original configuration
space to a new manifold $\gh M$, locally coordinated by a new set of
fields $\Phi^A$, $A=1,\ldots, N,$ (including, apart from the original
fields, ghosts, antighosts, etc.) and
their associated antifields $\Phi^*_A$, with opposite
Grassmann parity. This set
is often collectively denoted by
$z^a=\{\Phi^A, \Phi^*_A\}$, $a=1,\ldots, 2N$.
Afterwards, $\gh M$ is endowed with an odd symplectic
structure, $(\cdot,\cdot)$, called antibracket and defined as
$$
 (X,Y) = \frac{\partial_r X}{\partial z^a } \zeta^{ab}
         \frac{\partial_l Y}{\partial z^b }
\ , \quad \quad {\rm where } \quad
  \zeta^{ab} \equiv (z^a, z^b)=
  \left(
  \begin{array}{cc}
   0 & \delta^A_B\\
   -\delta^A_B & 0
  \end{array}\right).
$$

At the classical level, the fundamental object
is a bosonic functional $S(z)$ with dimensions of action and
verifying the so-called classical master equation,
\be
   (S,S)=0,
\label{cme}
\ee
with boundary conditions:
1) Classical limit:
$\restric{S(\Phi,\Phi^*)}{\Phi^*=0}= S_0(\phi)$, and
2) Properness condition:
${\rm rank}\restric{(S_{ab})}{\rm on-shell}=N,$
with
$ S_{ab}\equiv
  \left(\frac{\partial_l\partial_r S}{\partial z^a\partial z^b}\right),
$
and where on-shell means on the
surface $\left\{\frac{\partial_r S}{\partial z^a}=0\right\}$.

In the original basis $z^a$, the expansion of $S$
in antifields
\be
  S(\Phi,\Phi^*)=S_0(\phi)+\Phi^*_A R^A(\Phi)
  +\frac12\Phi^*_A\Phi^*_B R^{BA}(\Phi)+\ldots,
\label{scb}
\ee
generates the structure
functions of the original classical gauge algebra \cite{DWVH}. Besides,
fulfillment
of eq.\bref{cme} provides the relations defining its structure
\cite{BV85} \cite{FH90}.
In this sense, it is sensible to call the original basis $z^a$
{\it classical basis}%
\footnote{This definition of classical basis
differs from that given in \cite{VP93} \cite{TP93}. There,
this concept is based in the ghost number carried by
fields and antifields, while in this approach it lies on what it is
obtained in the $\Phi^*=0$ (classical) limit and in the content of
$(S,S)=0$.} \cite{VP93} \cite{TP93}.

The gauge-fixed theory, instead, is better
analyzed in terms of what can be called
{\it gauge-fixed basis}%
\footnote{The concept of gauge-fixed basis used throughout this paper,
based in \bref{gf}, in which fields
$\Phi^A$ do not change, is more restrictive than
that considered in \cite{VP93} \cite{TP93}, where
interchange of fields and antifields is also allowed.}
\cite{VP93} \cite{TP93},
defined in terms of a
suitable gauge-fixing fermion $\Psi(\Phi)$ through the
canonical transformation \cite{BV81}
\be
  \Phi^A\rightarrow \Phi^A,\quad\quad
  \Phi^*_A\rightarrow K_A+\frac{\partial \Psi(\Phi)}{\partial \Phi^A}
  \equiv K_A+\Psi_A.
\label{gf}
\ee
Then, in the gauge-fixed basis, ${z'}^a=\{\Phi^A, K_A\}$,
the proper solution $S$ \bref{scb}
is expressed as
\bea
  &&\hat S(\Phi,K)\equiv
  S\left(\Phi,\Phi^*=K+\frac{\partial \Psi(\Phi)}{\partial \Phi}\right)=
\nonumber\\
  &&\left(S_0(\phi)+
  \Psi_A R^A +\frac12 \Psi_A \Psi_B R^{BA}+\ldots\right)
  +K_A \left( R^A + \Psi_B R^{BA}+\ldots\right)+\gh O(K^2).
\label{sgf0}
\eea

The antifield independent part of \bref{sgf0} is the
gauge-fixed action, $S_\Sigma(\Phi)$,
so that the classical limit is no longer recovered.
This result provides
the characterization of the gauge-fixed
basis, through the boundary conditions:
1') Gauge-fixed limit:
$\restric{\hat S(\Phi, K)}{K=0}= S_{\Sigma}(\Phi),$ and
2')
${\rm rank}\restric{(S_{\Sigma,AB})}{\rm on-shell}=N,$
with
$ S_{\Sigma,AB}\equiv
 \left(\frac{\partial_l\partial_r S_\Sigma(\Phi)}
 {\partial \Phi^A\partial \Phi^B}\right ).
$
Therefore, if $\Psi$ in \bref{gf} is
correctly chosen \cite{BV81},
propagators are well defined and the usual perturbation theory can comence.

In much the same way, the linear part in $K_A$ of \bref{sgf0}
is the gauge-fixed BRST transformation of $S_\Sigma(\Phi)$
\be
  \delta_\Sigma\Phi^A= \restric{(\Phi^A, \hat S)}{K=0}\equiv\tilde R^A,
\label{gfbrst}
\ee
the coefficients of the bilinear part are the non-nilpotency
structure functions and so on. We can write
\be
  \hat S(\Phi,K)=S_\Sigma(\Phi)+K_A \tilde R^A(\Phi)
  +\frac12 K_A K_B \tilde R^{BA}(\Phi)+\ldots.
\label{exp s}
\ee
in such a way that $\hat S$ \bref{exp s} appears now as the
generating functional of the structure functions which define the BRST
symmetry, whereas relations derived
from $(\hat S,\hat S)=0$ characterize
the structure of the classical BRST symmetry.
\bref{exp s} contains thus all the information about
the underlying structure of this classical BRST symmetry.

$\hat S$ \bref{exp s} is itself invariant under the
off-shell nilpotent BRST symmetry
\be
  \delta F(z')= (F,\hat S).
\label{ngfbrst}
\ee
The cohomology associated with $\delta$, usually called
antibracket BRST cohomology, is related with the weak cohomology of
$\delta_\Sigma$ \bref{gfbrst} \cite{FH90}. Both cohomologies
turns out to be very important in the study of
renormalization and anomaly issues.

\subsection{Quantum theory. Quantum BRST transformation}

\hspace{\parindent}

The transition from the classical to the quantum theory
may spoil the classical BRST structure due to
quantum corrections
acquired by the BRST transformations \bref{gfbrst} and the higher
order structure functions in \bref{exp s}.
This violation indicates the presence of anomalies.
The quantum aspects of the BRST formalism are most suitable studied in
terms of the effective action $\Gamma$
associated through a Legendre type transformation
with respect to the sources $J_A$ with the (connected part of)
the generating functional
\be
  Z(J,K)=\int\gh D \Phi\exp\left\{\frac{i}{\hbar}
  \left[W(\Phi,K)+J_A\Phi^A\right]\right\},
\label{gen func}
\ee
with $W(\Phi,K)$ given by
\be
W(\Phi,K)=\hat{S}(\Phi,K)
+\sum^\infty_{p=1}{\hbar}^p M_p(\Phi,K),
\label{exp w}
\ee
and where the local counterterms $M_p$
should guarantee the finiteness of the theory while preserving the
BRST structure at quantum level as far as possible.
In this way, the functional $\Gamma$ appears as the quantum analog of
$\hat S$ \bref{exp s}, i.e., the coefficients in its antifield expansion%
\footnote{The effective action $\Gamma$ depends in fact on the
so-called ``classical fields'',
$\Phi^A_c(J,K)=-i\hbar\lder{\ln Z(J,K)}{J_A}$.
However, for notational simplicity, and unless confusion arise, we will
denote them also by $\Phi^A$.}
\be
  \Gamma(\Phi,K)=\Gamma(\Phi)+K_A \Gamma^A(\Phi)
  +\frac12 K_A K_B \Gamma^{BA}(\Phi)+\ldots,
\label{exp gamma}
\ee
are interpreted as the quantum counterpart of the classical coefficients in
$\hat S$: $\Gamma(\Phi)$ is the
1PI generating functional for the basic fields, including
loop corrections to $S_\Sigma(\Phi)$;
$\Gamma^A(\Phi)$
the quantum BRST transformations,
formed by adding quantum corrections to $\tilde R^A$ \bref{gfbrst};
$\Gamma^{AB}(\Phi)$ the quantum non-nilpotency structure
functions, etc. $\Gamma(\Phi, K)$ is thus the generating functional of the
structure functions characterizing the BRST symmetry at quantum level.

The quantum BRST structure and its possible violation are reflected in the
(anomalous) BRST Ward identity
\be
   \frac12(\Gamma,\Gamma)=-i\hbar(\gh A\cdot\Gamma),\quad\quad
   \gh A\equiv
   \left[\Delta W+\frac{i}{2\hbar}(W,W)\right](\Phi,K),
\label{wi}
\ee
where $(\gh A\cdot\Gamma)$ denotes the generating functional of
the 1PI diagrams with the insertion of
$\gh A$ and $\Delta$ stands for the second order differential operator
$$
   \Delta\equiv (-1)^A
   \frac{\partial_l}{\partial\Phi^A}
   \frac{\partial_l}{\partial\Phi^*_A}.
$$
Therefore, $\gh A$ in \bref{wi} parametrizes potential departures from
the classical BRST structure due to quantum corrections. In particular,
in its $K_A$ expansion,
the antifield independent part
indicates the non-invariance of
$\Gamma(\Phi)$ under the quantum BRST transformation $\Gamma^A(\Phi)$, its
linear part in $K_A$ reflects an anomaly in the
on-shell nilpotency of
$\Gamma^A(\Phi)$, and so on.

Quantum BRST invariance will thus hold if the obstruction
$\gh A$
in \bref{wi} vanishes, i.e., upon fulfillment through a local object $W$ of
the quantum master equation
\be
   \frac12(W,W)-i\hbar\Delta W=0.
\label{qme}
\ee
Eq.\bref{qme} encodes at once the classical master equation
\bref{cme}, satisfied by construction, plus a set of recurrent
equations for the counterterms $M_p$
\bea
     (M_1,\hat S)&=& i\Delta \hat S,
\label{first order}\\
     (M_p,\hat S)&=& i\Delta M_{p-1}-\frac12\sum^{p-1}_{q=1}
     (M_q,M_{p-q}),\quad\quad p\geq 2.
\nonumber
\eea

\subsection{Regularized FA formalism. Pauli-Villars scheme}

\hspace{\parindent}%
To make sense out of the previous calculations and expressions
a regularization scheme is necessary.
A prescription to regularize the FA formalism up to one loop
has been considered in refs.\cite{TNP89} \cite{T1}
\cite{TP93} \cite{Frank}. This proposal consists in using a
Pauli-Villars (PV) regularization
scheme in \bref{gen func}, that is, to substitute
it for the one-loop regularized expression
\be
  Z_{\rm reg}(J,K)=\int\gh D \Phi \gh D \chi
  \exp\left\{\frac{i}{\hbar}
  \left[\hat S(\Phi,K)+\hbar M_1(\Phi, K)
  + S_{\rm PV}(\chi,\chi^*=0;\Phi, K)+J_A\Phi^A\right]\right\},
\label{reg gen func}
\ee
where the PV fields $\chi^A$, introduced for each field $\Phi^A$, have the
same statistics as their original partners, but with the
path integral formally defined so that an extra minus sign occurs in
front of their loops.
Each PV field $\chi^A$ comes with its associated antifield
$\chi^*_A$, and together can collectively be denoted as
$w^a=\{\chi^A, \chi^*_A\}$, $a=1,\ldots,2N$. PV antifields
$\chi^*_A$ have no physical significance and at the end are put to zero.
Finally, the regularized theory is
described by \bref{reg gen func}
when the cutoff mass $M$ is sent to infinity.

The regulating PV action $S_{\rm PV}$ is determined from
two requirements: i) massless propagators
and couplings for PV fields should coincide
with those of their partners
and ii) BRST transformations for PV fields should
be such that the masless part of the PV action,
$S_{\rm PV}^{(0)}$, and the measure in
\bref{reg gen func} be BRST invariant up
to one loop. A suitable prescription for $S_{\rm PV}$
is \cite{T1} \cite{TP93}
\be
  S_{\rm PV}= S^{(0)}_{\rm PV}+S_M=
    \frac12 w^a S_{ab} w^b
    -\frac12 M \chi^A T_{AB} \chi^B,
\label{orig pv action}
\ee
with the mass matrix $T_{AB}$ arbitrary but invertible and $S_{ab}$
defined by
\be
    S_{ab}=\left(\lder{}{{z'}^a}\rder{}{{z'}^b}\hat{S}(\Phi, K)\right).
\label{s der}
\ee

Let us now derive an alternative form for the regularized expression of
$\Delta \hat S$. Application of the
semiclassical approximation to \bref{reg gen func}
yields the effective action up to one loop
\be
   \Gamma(\Phi, K)= \hat S(\Phi, K)+\hbar M_1(\Phi,K)+
   \frac{i\hbar}2{\rm Tr}\ln\left[\frac{(TR)}{(TR)-TM}\right]=
   \hat S(\Phi, K)+\hbar \Gamma_1(\Phi,K),
\label{gamma1}
\ee
where Tr stands for the supertrace,
${\rm Tr}(M)\equiv [(-1)^A M^A_{\;A}]$, and
$(TR)_{AB}$ is defined from \bref{s der} as
\be
   (TR)_{AB}=\left(\frac{\partial_l}{\partial\Phi^A}
   \frac{\partial_r}{\partial\Phi^B} \hat{S}(\Phi, K)\right).
\label{orig pv kin}
\ee
On the other hand, the BRST variation
\bref{ngfbrst} of $\Gamma_1$ in \bref{gamma1}
produces, by comparison with \bref{wi}, what should be considered
the regularized expression of $\Delta \hat S$
\be
   (\Delta \hat S)_{\rm reg}=
   \delta\left\{-\frac12{\rm Tr}\ln\left[\frac{R}{R-M}\right]\right\}=
   {\rm Tr}\left[-\frac12(R^{-1}\delta R)
   \frac{1}{(1-R/M)}\right].
\label{a 2 quad}
\ee
In appendix A, we prove the equivalence between
\bref{a 2 quad} and the form obtained in \cite{T1} \cite{TP93}.

The expression we have obtained for the regularized
value of $\Delta \hat S$ involves the
BRST variation of the regulator, $\delta R$, thereby showing that
(potential) anomalous symmetries are
directly related with the transformation properties of $R$.
In particular, if $R$ is invariant under some subset of
symmetries or it transforms as $\delta R=[R, G]$ for a given $G$,
\bref{a 2 quad} leads to a vanishing result.
On the other hand, since $(\Delta \hat S)_{\rm reg}$ appears also
as a $\delta$-variation, its BRST variation itself vanishes,
$\delta\left[(\Delta \hat S)_{\rm reg}\right]=0$,
i.e., it verifies the Wess-Zumino consistency conditions \cite{WZ71}%
\footnote{For an alternative proof of the consistency of
$(\Delta \hat S)_{\rm reg}$, see \cite{VP93}.}.
Therefore, the complete expression of $(\Delta \hat S)_{\rm reg}$
that is, the ``anomaly'' (term
in $M^0$) plus the divergent terms (terms in $M^n$, $n>0$) verify
separately the Wess-Zumino consistency conditions.

Finally, and basically for computational reasons, it should be stressed
that $\Delta \hat S$ \bref{a 2 quad} is
equivalent \cite{TNP89}
to that obtained using the well known Fujikawa
regularisation procedure \cite{Fuj80} \cite{Fuji}. In other words
$$
   (\Delta \hat S)_{\rm reg}=
   {\rm Tr}\left[-\frac12(R^{-1}\delta R)
   \frac{1}{(1-R/M)}\right]
   \sim
   {\rm Tr}\left[-\frac12(R^{-1}\delta R)
   \exp\{R/ M\}\right].
$$

\section{Analysis of $(\Delta \hat S)_{\rm reg}$}

\hspace{\parindent}%
The aim of this section is to present a detailed analysis of the
regularized expression of $\Delta \hat S$ \bref{a 2 quad} arising in the
PV regularization scheme. This study is based on some results
about the so-called antibracket BRST cohomology associated with the
BRST operator $\delta$ \bref{ngfbrst} and its relation with the weak
cohomology of $\delta_\Sigma$ \bref{gfbrst}, which can be found in
\cite{FH90}.

Consider the regularized expression of $\Delta \hat S$ \bref{a 2 quad}.
Without loss of generality, we can assume for $R$ an
expansion in antifields of the type
\be
  R(\Phi,K)=R_0(\Phi)+K_A R^A+\gh O(K^2),
\label{exp r}
\ee
with $R_0(\Phi)$ invertible.
$\delta R$ becomes then in terms of $\delta_\Sigma$
\bref{gfbrst}
$$
  \delta R(\Phi, K) =\delta_\Sigma R_0(\Phi)
  +(-1)^R\frac{\partial_r S_\Sigma}{\partial\Phi^A} R^A
  +\gh O(K).
$$
Now, upon substitution of the above expansions in \bref{a 2 quad},
$(\Delta \hat S)_{\rm reg}$ acquires the form
\be
   (\Delta \hat S)_{\rm reg}
  = \left({\rm Tr}\left[-\frac12(R_0^{-1}\delta_\Sigma R_0)
   \frac{1}{(1-R_0/M)}\right]
  -\frac{\partial_r S_\Sigma}{\partial\Phi^A} P^A\right)(\Phi)
  +\gh O(K).
\label{delta s anti}
\ee

The coefficients $P^A(\Phi)$ can be shown to be local under certain
conditions, in the same way as locality of
\bref{a 2 quad} (or even of
the trace term in \bref{delta s anti}) is proven%
\footnote{We are indebted to A. van Proeyen and W. Troost for clarifying
this point to us.} \cite{Gil}.
In this case, the combination
of the equations of motion in \bref{delta s anti}
can be expressed in a local way as
\be
  -\frac{\partial_r S_\Sigma}{\partial\Phi^A} P^A(\Phi)
  +\gh O(K)= (K_A P^A, \hat S)+\gh O(K).
\label{pa}
\ee
On the other hand, by comparison with \bref{a 2 quad},
the trace term in \bref{delta s anti} can also be written as
\be
   {\rm Tr}\left[-\frac12(R_0^{-1}\delta_\Sigma R_0)
   \frac{1}{(1-R_0/M)}\right]= \delta_\Sigma
  \left\{-\frac12{\rm Tr}\ln\left[\frac{R_0}{R_0-M}\right]\right\}
   \equiv ({\Delta \hat S})^{(0)}_{\rm reg}(\Phi).
\label{tilde delta s zero}
\ee
Collecting thus the above results \bref{pa} and \bref{tilde delta s zero},
and absorbing the dependence on the (BRST trivial) auxiliar sector
of fields through a local counterterm $\gh N$, we can write
\be
   (\Delta \hat S)_{\rm reg}=
    \gh B(\Phi_{\rm m})+ (\widehat{\Delta \hat S})_{\rm reg}
    + (\gh M,\hat S),
\label{result}
\ee
with $(\widehat{\Delta \hat S})_{\rm reg}\sim \gh O(K)$,
$\gh M=K_A P^A+\gh N$, $\{\Phi_{\rm m}\}$ the minimal sector of
fields and $\gh B(\Phi_{\rm m})$ the projection to this minimal sector
of $(\Delta \hat S)^{(0)}_{\rm reg}(\Phi)$ \bref{tilde delta s zero}, i.e.,
\be
  \gh B(\Phi_{\rm m})\equiv
  \restric{
  (\Delta \hat S)^{(0)}_{\rm reg}}{\Phi_{\rm m}}=
   \delta_\Sigma
   \left\{-\frac12{\rm Tr}\ln\left[\frac{R'_0}{R'_0-M}\right]\right\},
   \quad\quad\mbox{\rm with}\quad\quad R'_0\equiv
  \restric{R_0}{\Phi_{\rm m}}.
\label{ghb}
\ee

The result \bref{result}
has some relevant consequences when it is applied to "closed"
theories. As that, we mean theories for which $\hat S$ \bref{exp s} is at
most linear in the antifields, implying that
$\delta_\Sigma\Phi=\delta\Phi$ and
$\delta^2_\Sigma=0$ ``strongly'', i.e., without use of the equations of
motion. Indeed, in such cases, \bref{ghb} and, as a consequence,
$(\widehat{\Delta \hat S})_{\rm reg}$ in \bref{result}, are seen to
satisfy
$\delta \left[ \gh B(\Phi_{\rm m}) \right]=
  \delta_\Sigma \left[ \gh B(\Phi_{\rm m}) \right]=
  \delta_\Sigma^2\{\cdot\}=0$ and
$\delta[(\widehat{\Delta \hat S})_{\rm reg}]=0$ {\it separately}.
One would then be tempted to use the isomorphism between
$\delta$ and $\delta_\Sigma$ cohomologies
stated in \cite{FH90} and write
$(\widehat{\Delta \hat S})_{\rm reg}$ in \bref{result}
as $\delta[\tilde{\gh O}(K)]$. However,
as one can infer from the analysis in \cite{VP93},
this isomorphism holds for local functions,
but not in general for integrals of local functions.
A priori, thus, we can not
conclude the $\delta$ triviality of $(\widehat{\Delta \hat S})_{\rm reg}$,
which should be studied in principle case by case.

However, two important facts can be derived from the above analysis when
dealing with closed theories:

\begin{itemize}
\item
The antifield independent part of
\bref{result} can be cohomologically studied separately
from that containing antifields, $(\widehat{\Delta \hat S})_{\rm reg}$.
In this way, while
$\gh B(\Phi_{\rm m})$ expresses the non BRST invariance of
$\Gamma(\Phi)$ in \bref{exp gamma} at one loop,
$(\widehat{\Delta \hat S})_{\rm reg}$
is likely to be related to anomalies in the nilpotency of the
quantum BRST transformation and in higher order relations which define
the BRST structure. Such kind of terms may be of interest for theories
for which the BRST charge suffers from anomalies at quantum level (bosonic
string, etc.). Their study, however, goes beyond the scope of this paper
and from now on we will restrict ourselves to the analysis of the antifield
independent part.
\item
\bref{result} indicates that
anomalies with non-trivial antifield dependence of the type presented in
\cite{Brandt} are not expected to appear in this formulation.
The reason behind is that they require
the presence of an antifield independent part $\delta_\Sigma$-invariant
{\it on-shell}, while in the above regularization procedure, and under the
above assumed and plausible locality of the coefficients $P^A$,
the antifield independent part of \bref{result}, $\gh B(\Phi_{\rm m})$,
results to be {\it off-shell} $\delta_\Sigma$ (or $\delta$) invariant.
Therefore, it is as if the regularization procedure selects
from the complete set of ghost number one non-trivial
cocycles a subset of ``physical anomalies'', i.e., of candidates to be
realized in the given theory. In any case, this mismatch between
mathematical solutions and ``physically'' realized solutions deserves
further investigation.
\end{itemize}

\section{Regularised Field-Antifield formalism for anomalous gauge
         theories}

\subsection{General discussion}

\hspace{\parindent}%
Let us consider an irreducible theory with closed algebra.
The minimal sector of
fields consists then of the classical fields $\phi^i$ and the ghosts
$\gh C^\alpha$. Under such conditions, the regulator $R$
\bref{exp r} could be written in fact as
\be
  R(\Phi,K)=\gh R(\phi)+\widehat R(\Phi,K),
\label{exp r min}
\ee
with $\gh R(\phi)$ invertible, so that \bref{ghb} becomes,
in comparison with \bref{a 2 quad}
\be
  \gh B(\Phi_{\rm m})=
  {\rm Tr}\left[-\frac12(\gh R^{-1}\delta\gh R)
   \frac{1}{(1-\gh R/M)}\right]= A_\alpha(\phi)\gh C^\alpha.
\label{a alpha}
\ee

Assume now that no
local counterterm $M_1(\Phi, K)$
exists satisfying \bref{first order}, or equivalently, that no local
counterterms $M_1^{(0)}(\phi)$,
$M'_1(\Phi,K)\sim\gh O(K)$, exist satisfying
\bea
   (M_1^{(0)}, \hat S)&=&i A_\alpha(\phi)\gh C^\alpha=
   i a_k A^k_\alpha(\phi)\gh C^\alpha,
\label{47}\\
   (M'_1,\hat S)&=& i (\widehat{\Delta \hat S})_{\rm reg},
\label{48}
\eea
with $\{A^k_\alpha(\phi)\}$ a basis of BRST
nontrivial cocycles at ghost number one and where $A_\alpha(\phi)$, from
now on, will only stand for the finite pieces in \bref{a alpha} (i.e.,
divergent pieces are assumed to be
absorbed by the BRST variation of local terms to be included in
$M_1$ in \bref{exp w}.).
We will restrict ourselves to the study of \bref{47}, postponing the
analysis of antifield dependent issues, as \bref{48}, to the future.

The rank of the functional derivatives of the anomalies $A_\alpha(\phi)$
\be
  {\rm rank}
  \left(\frac{\partial A_\alpha(\phi)}{\partial\phi^i}\right)=r(\leq m),
  \quad\quad\alpha=1,\ldots,m,
\label{rank a}
\ee
determines the number of anomalous gauge transformations and, as a
byproduct, the number of pure gauge degrees of freedom that become
propagating at quantum level. In the case
$r<m$, it often happens that the original gauge transformations
\bref{orig gau} split as
$$
  \delta\phi^i=R^i_\alpha\veps^\alpha=
  R^i_A\veps^A+ R^i_a\veps^a=\delta_{(A)}\phi^i+\delta_{(a)}\phi^i,
  \quad\quad a=1,\ldots,r<m,
$$
in such a way that the regulator $\gh R(\phi)$ in
\bref{exp r min} ``preserves'' the $A$ part, i.e.,
\be
   \delta_{(A)} \gh R=
   \left\{\begin{array}{l}
    0  \\
    {[\gh R, G_A\veps^A]}
   \end{array}\right.,
   \quad\quad\mbox{but}\quad
   \delta_{(a)} \gh R\neq
   \left\{\begin{array}{l}
    0  \\
    {[\gh R, G_a\veps^a]}
   \end{array}\right. .
\label{deltas r 1}
\ee
Then, \bref{a alpha} yields $\gh B(\Phi_{\rm m})= A_a(\phi)\gh C^a$,
where $A_a(\phi)$ are assumed to be independent.
This situation is only possible when the $A$
part is a subgroup \cite{GP93}, while no restrictions exist for the $a$
part. For the sake of simplicity, in the rest of the paper
we consider the $a$ part to be also a subgroup.

Let us sketch now the general ideas of the extended formalism in the case
$r=m$ \cite{GP93}. The proposal consists
in introducing $r=m$ new fields $\theta^\alpha$, and demand
that their gauge transformations
\be
  \delta\theta^\alpha=-\tilde\mu^\alpha_\beta(\theta,\phi) \veps^\beta,
\label{extra trans}
\ee
are such that $\phi^a=(\phi^i,\theta^\alpha)$%
\footnote{For simplicity, we will consider only bosonic fields and bosonic
gauge transformations, i.e., $\eps(\phi^i)=0$, $\eps(\veps^\alpha)=0$.}
constitutes a new
representation of the original gauge group. In terms of the generator
$R^a_\alpha=(R^i_\alpha,-\tilde\mu^\beta_\alpha)$,
this requirement amounts to the condition that
$R^a_\alpha$ verifies the same algebra as the original ones
$R^i_\alpha$ \bref{orig gau}. An explicit solution is \cite{GP93}
\be
   \tilde\mu^\alpha_\beta(\theta,\phi)=\restric{
   \frac{\partial\phi^\alpha(\theta',\theta;\phi)}
   {\partial\theta^{'\beta}}}{\theta'=0},
\label{mu tilde}
\ee
$\phi^\alpha(\theta',\theta;\phi)$ being the composition functions of
the gauge (quasi)group%
\footnote{For a complete study of the so-called quasigroup structure, as
well as for further explanation of notation, we refer the reader to the
original reference \cite{B81}.}.
In this way, the extension enlarges
the classical physical content of the extended theory.
Indeed, the finite gauge
transformations of the classical fields with parameters $\theta^\alpha$,
$F^i(\phi,\theta)$, are gauge invariant,
$\delta F^i(\phi,\theta)=0$, and can thus be
considered $n$ new classical gauge invariant degrees of freedom of the
extended theory. The extension procedure is then prepared
to describe already at classical level the (quantum) appearence of new
degrees of freedom.

In the FA framework, all these facts are summarized
in the (non-proper)
solution of the classical master equation in the extended space \cite{GP93}
\be
  S_{\rm ext}=S- \theta^*_\alpha\tilde\mu^\alpha_\beta\gh C^\beta\equiv
  S+S_\theta,
\label{ext sol}
\ee
where $S$ is the original proper solution and $\theta^*_\alpha$ the
antifields associated with the extra fields $\theta^\alpha$.
Its non-proper character \cite{hid}
is just a consequence of the absence of terms related with the shift
symmetry $\delta\phi^i=0$, $\delta\theta^\alpha=\sigma^\alpha$,
of the classical action $S_0(\phi)$.

To quantize the extended theory, consider an extension
of $W$ \bref{exp w} at one loop%
\footnote{The gauge-fixed basis we consider in the extended theory come
from gauge-fixing fermions of the type $\Psi(\Phi)$. This implies that
neither $\theta$ nor $\theta^*$ change under \bref{gf}.}
\be
   \tilde W= \hat{S}_{\rm ext}+\hbar \tilde M_1,
\label{tilde w}
\ee
verifying \bref{cme} and \bref{first order}
in the extended space
\bea
     ( \hat S_{\rm ext}, \hat S_{\rm ext})&=&0,
\label{cme2}\\
     (\tilde M_1, \hat S_{\rm ext})&=& i\Delta \hat S_{\rm ext}.
\label{new first order}
\eea

In principle, one would be tempted to look for a proper solution
$S_{\rm ext}$ of \bref{cme2} (as in ref.\cite{hid}), expressed in the
gauge-fixed basis. However, since pure gauge degrees of freedom become
propagating due to quantum corrections, we claim that in this proposal
the classical part of $\tilde W$ \bref{tilde w}, $\hat S_{\rm ext}$, should
describe only the propagation of the original fields $\Phi^A$,
while propagation of the extra fields $\theta^\alpha$
should be provided by the
first quantum correction $\tilde M_1$. In other words,
using the collective notation
$\Phi^{\mu}=\{\Phi^A,\theta^\alpha\}$,
we are led to use
a solution $\hat S_{\rm ext}$ of \bref{cme2} for which
\be
 {\rm rank}\restric{(\hat S_{{\rm ext},\mu\nu})}{\rm on-shell}\equiv
 {\rm rank}\restric
 {\left(\frac{\partial_l\partial_r  \hat S_{\rm ext}}
 {\partial \Phi^{\mu}\partial \Phi^{\nu}}\right )}
 {\rm on-shell}=N,
\label{rank cond}
\ee
whereas $\tilde W$ \bref{tilde w} should verify
\be
 {\rm rank}\restric{(\tilde W_{\mu\nu})}{\rm on-shell}\equiv
 {\rm rank}\restric
 {\left(\frac{\partial_l\partial_r \tilde W}
 {\partial \Phi^{\mu}\partial \Phi^{\nu}}\right )}
 {\rm on-shell}=N+m.
\label{rank cond 2}
\ee

A convenient solution of \bref{cme2} turns out to be
\bref{ext sol} \cite{GP93}. Indeed, it is
non-proper, whereas
\bref{rank cond} is guaranteed since $\hat S_{\rm ext}$ contains the
original proper solution $\hat S$. With respect to
the quantum corrections,
eq.\bref{new first order} is (formally)
specified once $\Delta \hat S_{\rm ext}$ is known. A formal computation
\be
   \Delta \hat S_{\rm ext}=
   \Delta \hat S-\tilde\mu^\beta_{\alpha,\beta}\gh C^\alpha=
   \Delta \hat S+\Delta S_\theta,
\label{delta t s}
\ee
indicates that the new degrees of freedom modifies
$\Delta \hat S$. The new contribution
is the unregularized logarithm of the jacobian of the
BRST transformation for the $\theta^\alpha$ fields. This fact indicates
that the regularization procedure
should be adapted to the extended theory in order to
take into account contributions coming from the extra degrees of freedom.

However, the PV regularization program can not be applied in a direct way,
basically because
\bref{rank cond} implies that a
``kinetic term'' for $\theta^\alpha$ is lacking in
$\hat S_{\rm ext}$.
This drawback can be bypassed by taking the
ansatz for the quantum action $\tilde W$ \bref{tilde w}
$$
  \tilde W= [\hat S_{\rm ext}+\hbar M^{(0)}_1]+\hbar M',
$$
with $M'$ containing the original one $\gh M$
in \bref{result} (and possibly including $\theta$ dependent terms
at least of $\gh O(K)$ solving \bref{48} in a local way) and where%
\footnote{$\lambda^\alpha_\beta$ in \bref{wz} is the inverse matrix of
$\mu^\alpha_\beta=\restric{
   \frac{\partial\phi^\alpha(\theta,\theta';\phi)}
   {\partial\theta^{'\beta}}}{\theta'=0}$.}
\be
   M^{(0)}_1(\phi,\theta)=-i\int_0^1 A_\beta
   (F(\phi,\theta t))\lambda^\beta_\alpha(\theta t,\phi)
   \theta^\alpha \dif t,
\label{wz}
\ee
is the original Wess-Zumino term \cite{WZ71}, i.e, the solution in the
extended space of \cite{GP93}
\be
  (M^{(0)}_1, \hat S_{\rm ext})= i A_\alpha(\phi)\gh C^\alpha.
\label{eqwz}
\ee
Indeed, with this choice, $\tilde W_{\mu\nu}$
contains now basically the original hessian $\hat S_{AB}$ plus a new
nondiagonal block for the extra variables $\theta^\alpha$, which
essentially reads
$$
   \left(\frac{\partial^2 M^{(0)}_1(\phi,\theta)}
       {\partial\phi^i\partial\theta^\alpha}\right)=-i
   \left(\frac{\partial A_\alpha(\phi)}{\partial\phi^i}\right)
           +\gh O(\theta).
$$
In this way, taking into account \bref{rank a},
this ansatz gives the correct rank \bref{rank cond 2}
for $\tilde W_{\mu\nu}$.
Summarizing, at this first stage it seems plausible to consider
as the action to regularize the extended theory
\be
  S'=[\hat S_{\rm ext}+\hbar M^{(0)}_1].
\label{new s}
\ee

However, altough this ansatz solves the first problem, some others appear:
\begin{itemize}
\item i) The new action $S'$ \bref{new s}
does not verify the classical master equation
\be
  (S',S')= 2 i \hbar A_\alpha(\phi)\gh C^\alpha\neq 0,
\label{sprim2}
\ee
so that the PV procedure of sect.2 can no longer be considered.

\item ii) The part providing for the propagation of the $\theta^\alpha$
fields in \bref{new s} contains explicitly an $\hbar$%
\footnote{In the effective theories of the standard model, where some
heavy fermions are integrated out, one should also consider Wess-Zumino
terms to take into account the presence of the anomaly. In this case,
however, the extra variables are present already in the classical action
and the above difficulties for the propagator of the extra varibles do not
appear \cite{Fer}.}.
This fact would ruin
the usual $\hbar$ perturbative expansion and the
tool to recognize one-loop anomalies.

\end{itemize}
A sensible PV regularization of the extended theory requires thus
to extract from $S'$ \bref{new s}
a classical part $W_0$ which constitutes a proper
solution in the extended space, that is, satisfying
\begin{itemize}
\item  a): $( W_0, W_0)=0,$
\item  b):
$ {\rm rank}\restric{(W_{0,\mu\nu})}{\rm on-shell}\equiv
 {\rm rank}\displaystyle{\restric
 {\left(\frac{\partial_l\partial_r W_0}
 {\partial \Phi^{\mu}\partial \Phi^{\nu}}\right )}
 {\rm on-shell}}=N+m. $
\end{itemize}
In what follows, we will see how this splitting can be implemented
for certain systems through canonical transformations in the
extra variables sector.

\subsection{The extended proper solution $W_0$. Background terms}

\hspace{\parindent}%
In order to see how to get $W_0$,
let us analyze the $\theta^\alpha$
and $\theta^*_\alpha$ dependent parts of $S'$ \bref{new s} by expanding
them in powers of $\theta^\alpha$. Working in a canonical
parametrization
( $\lambda^\alpha_\beta\theta^\beta=\theta^\alpha$)
\cite{B81}, we have for \bref{wz}
\bea
   \hbar M^{(0)}_1(\phi,\theta)&=&
   -i\hbar\left[A_\alpha(\phi)\theta^\alpha
   +\frac12 \theta^\alpha D_{\alpha\beta}(\phi)\theta^\beta
   +\frac{1}{3!} \theta^\alpha\theta^\beta\theta^\gamma
   (\Gamma_\alpha D_{\beta\gamma})(\phi)\right.
\nonumber\\
   &&\hspace{9mm}\left.
   +\ldots
   +\frac{1}{n!} \theta^{\alpha_1}\ldots\theta^{\alpha_n}
   (\Gamma_{\alpha_1}\ldots \Gamma_{\alpha_{n-2}}
   D_{\alpha_{n-1}\alpha_n})(\phi)+\ldots\right],
\label{wz exp}
\eea
with $\Gamma_\alpha$ and $D_{\alpha\beta}$ defined by
\be
   \Gamma_\alpha=R^i_\alpha\frac{\partial}{\partial\phi^i},
   \quad\quad D_{\alpha\beta}=\Gamma_\beta A_\alpha=
   \left(\frac{\partial A_\alpha}{\partial\phi^i}R^i_\beta\right),
\label{dab}
\ee
while the $\theta^*_\alpha$ part in \bref{ext sol},
acquires the form
\be
 -\theta^*_\alpha\tilde\mu^\alpha_\beta\gh C^\beta=
 -\theta^*_\alpha
 \left[\delta^\alpha_\beta-
 \frac{1}{2} T^\alpha_{\beta\gamma}(\phi)\theta^\gamma
 +\gh O(\theta^2)\right] \gh C^\beta.
\label{exp mu}
\ee

It seems hence reasonable to make a redefinition of $\theta^\alpha$
such that it absorbs the $\hbar$ of their kinetic term,
and implement it to their antifields
through a canonical transformation \cite{hid}, i.e.,
\be
   \theta^{'\alpha}= \sqrt\hbar\,\theta^\alpha,\quad\quad
   \theta^{'*}_\alpha= \frac1{\sqrt\hbar}\,\theta^*_\alpha.
\label{teta trans}
\ee
In this way, expansions \bref{wz exp} and \bref{exp mu} become,
after dropping primes
\bea
   \hbar M^{(0)}_1(\phi,\theta)&\rightarrow&
   -i\left[\sqrt\hbar A_\alpha(\phi)\theta^\alpha
   +\frac{1}{2} \theta^\alpha D_{\alpha\beta}(\phi)\theta^\beta
   +\gh O(\theta^3; 1/\sqrt\hbar)\right],
\label{wz exp 2}\\
 -\theta^*_\alpha\tilde\mu^\alpha_\beta\gh C^\beta&\rightarrow&
 -\sqrt\hbar\theta^*_\alpha\gh C^\alpha
 +\frac12 \theta^*_\alpha T^\alpha_{\beta\gamma}(\phi)\theta^\gamma
 \gh C^\beta +\gh O(\theta^2;1/\sqrt\hbar),
\label{exp mu 2}
\eea
so that, although $\hbar$ dissapears in few terms or becomes
$\sqrt\hbar$, in higher order terms it appears in the
form of negative powers of $\sqrt\hbar$.
Therefore, it seems as if the quantum
treatment of Wess-Zumino terms can only be done in a nonperturbative
(in the $\hbar$ expansion sense) regime.
Whether or not this is true,
this perturbative treatment can at least be applied in a sensible way
to models for which only the first two terms
in \bref{wz exp 2} and \bref{exp mu 2} are really present.
In this case, the $\hbar^0$ terms
should be considered part of
$ W_0$, while the $\sqrt\hbar$ terms generalize
the so-called background charges \cite{Rom90}. In
the BV context these terms have been previously considered in \cite{hid}
\cite{VP93}. From now on, we will call them {\it background terms}.

Let us analyze the conditions that guarantee this perturbative treatment.
\bref{wz exp} stops at second order if
\be
     \Gamma_\gamma(D_{\alpha\beta})(\phi)=0.
\label{first cond}
\ee
On the other hand, the gauge transformations for
$\theta^\alpha$ would
read, in absence of $\gh O(\theta^2)$ terms
$$
  \delta\theta^\alpha=-\veps^\alpha+
  \frac12 T^\alpha_{\beta\gamma}(\phi)\theta^\gamma\veps^\beta,
$$
but in this form they can only provide a
representation of the original gauge algebra if
$T^\alpha_{\beta\gamma}=0$.
Summarizing, when $r=m$,
\bref{first cond} and
$T^\alpha_{\beta\gamma}=0$ are sufficient
conditions in order to have a sensible perturbative
expansion. In this case, after having performed
\bref{teta trans}, $S'$ \bref{new s} becomes
$$
  S'\rightarrow \left[\hat S(\Phi, K)
  -\frac{i}2 \theta^\alpha D_{\alpha\beta}(\phi)\theta^\beta\right]
  -\sqrt\hbar\left[
  \theta^*_\alpha\gh C^\alpha+i A_\alpha(\phi)\theta^\alpha\right]\equiv
  W_0+\sqrt\hbar M_{1/2},
$$
from which the form of $W_0$ can immediately be read off
\be
   W_0= \hat S(\Phi, K)
  -\frac{i}2 \theta^\alpha D_{\alpha\beta}(\phi)\theta^\beta.
\label{t w zero}
\ee

Under such conditions, eq.\bref{sprim2} translates to
\bea
  (W_0, W_0)&=&0,
\label{ncme}\\
  (W_0, M_{1/2})&=&0,
\label{inv back}\\
  \frac12 (M_{1/2},M_{1/2})&=& i A_\alpha(\phi)\gh C^\alpha,
\label{n anom}
\eea
thus indicating that $W_0$ \bref{t w zero} satisfies indeed
the classical master equation.

Finally, properness of $W_0$ holds if rank$(D_{\alpha\beta})=p=m$.
However, in view of
\bref{dab}, in general one can only guarantee $p\leq m$.
The extremum case appears, for example, when
the anomalies are gauge invariant, i.e.,
$D_{\alpha\beta}=0$ and $p=0$.
For the sake of brevity, from now on we will restrict
ourselves to the case $p={\rm max.}$, leaving the general case for the
future.

\subsection{Abelian anomalous subgroup}

\hspace{\parindent}
The above described situation is too restrictive
and somewhat trivial. In what follows, we consider the more
interesting case $r<m$ in \bref{rank a},
in which only an anomalous subgroup (the $a$ part) is abelian.

The extension procedure goes along the same lines discussed before
and is based in the introduccion of $r$ new fields $\theta^a$. The
subgroup character of the anomalous ($a$) part implies now
that the transformation for the new fields, the extended action
$S_{\rm ext}$ and the Wess-Zumino term $M^{(0)}_1$ are obtained from
\bref{extra trans}, \bref{ext sol} and \bref{wz} by simply considering the
$a$ subgroup as a group by itself; in brief, by the substitution
$\alpha\rightarrow a$ and by the restriction of all the quantities to
$\theta^A=0$.
The only non-trivial objects of this construction relevant for our
analysis are the type $A$ gauge generators
\be
  \tilde\mu^{'a}_B(\theta^a,\phi)=\restric{
  (\tilde\mu^a_B+\mu^a_b\lambda^b_D\tilde\mu^D_B)}{\theta^A=0},
\label{trans a 2}
\ee
with
$\tilde\mu^a_b$, $\tilde\mu^a_B$, $\mu^a_b$, $\lambda^b_D$ and
$\tilde\mu^D_B$ the corresponding blocks of the matrices
$\tilde\mu^\alpha_\beta$ \bref{mu tilde}, and $\mu^\alpha_\beta$,
$\lambda^\alpha_\beta$ defined above.
For a full description of the extension procedure in this case
we refer the reader to \cite{GP93}.

It is clear thus that, as far as the $a$ subgroup is concerned,
the previous conditions \bref{first cond} and $T^\alpha_{\beta\gamma}=0$
translates to
\be
     \Gamma_c(D_{ab})(\phi)=0, \quad\quad\mbox{\rm with}\quad\quad
     D_{ab}(\phi)=\Gamma_b A_a(\phi),
\label{first cond 2}
\ee
and $T^a_{bc}=0$, i.e., the $a$ subgroup should be abelian.
On the other hand, the expansion for the $A$
generators \bref{trans a 2} becomes, for
gauge invariant structure functions
$$
  \tilde\mu^{'a}_B(\theta^a,\phi)=\restric{
  (\tilde\mu^a_B+\mu^a_b\lambda^b_D\tilde\mu^D_B)}{\theta^A=0}=
  -T^a_{Bb}\theta^b
  +\frac14 T^a_{Dd} T^D_{Bb}\theta^b\theta^d
  +\frac1{24} T^a_{Dd} T^D_{Cc} T^C_{Bb}\theta^b\theta^c\theta^d
  +\ldots,
$$
thus indicating that the $A$ transformations
can be taken linear in $\theta^a$ if
\be
   T^a_{Dd} T^D_{Bb}=0.
\label{cond t}
\ee
This requirement is met, for instance, when either $T^a_{Dd}=0$ and/or
$T^D_{Bb}=0$. In the end, a direct computation shows that
$$
  \delta\theta^a=-\veps^a+ T^a_{Bb}(\phi)\theta^b\veps^B,
$$
provide a representation of the original gauge algebra if precisely
i) the structure functions are gauge invariant and ii) \bref{cond t} holds.
Summing up, conditions \bref{first cond 2}, \bref{cond t},
$T^a_{bc}=0$ and
$\Gamma_\sigma(T^\gamma_{\alpha\beta})=0$ guarantee
a sensible perturbative expansion for this extended theory.
Under such conditions,
the $A$ transformation for the kinetic operator $D_{ab}$
\bref{first cond 2} reads
\be
   \delta_{(A)} D_{ab}=
   \left(D_{ac} T^c_{bB}+ D_{bc} T^c_{aB}\right) \veps^B.
\label{dab trans}
\ee

Now, the canonical
transformation \bref{teta trans} adapted to this case
brings $S'$ \bref{new s} to the form
\be
  S'\rightarrow \left[\hat S(\Phi, K)
  -\frac{i}2 \theta^a D_{ab}(\phi)\theta^b
  +\theta^*_aT^a_{Bb}(\phi)\theta^b\gh C^B \right]
  -\sqrt\hbar\left[
  \theta^*_a\gh C^a+i A_a(\phi)\theta^a\right]\equiv
  W_0+\sqrt\hbar M_{1/2},
\label{wm12}
\ee
thus providing the following expression for $W_0$
\be
   W_0= \hat S(\Phi, K)
  -\frac{i}2 \theta^a D_{ab}(\phi)\theta^b
 +\theta^*_aT^a_{Bb}(\phi)\theta^b\gh C^B.
\label{t w zero 2}
\ee
Once again,
relations \bref{ncme}, \bref{inv back} and \bref{n anom} hold now for
$W_0$ and the background term
$M_{1/2}$. A direct check of these relations gives some
additional information. For example,
fulfillment of \bref{ncme} requires
$D_{ab}$ in \bref{t w zero 2}
to be the $a$ gauge variation of some consistent anomalies $A_a$,
while the vanishing of \bref{inv back},
instead, crucially relies on
$D_{ab}$ being precisely the $a$ gauge variation
of $A_a$ occurring in $M_{1/2}$.
These two conditions are summarized in the relation
$D_{ab}= \Gamma_b A_a$ in \bref{first cond 2}
Finally, the properness condition for $W_0$ holds as far as
rank$(D_{ab})=r$, which we will assume from now on.

\subsection{Regularization of the extended theory}

\hspace{\parindent}%
To develop the regularization procedure, we will be
mainly concerned with the last situation, since the first one
--once the substitution $a\rightarrow\alpha$ is
done-- can be seen as a special case of it,
with no antifields $\theta^*$ in $W_0$ \bref{t w zero 2}.

{}From the above discussion, and according to \bref{delta t s} and to
the form of $W_0$ \bref{t w zero 2}, it is expected that the new
variables generates an extra anomaly, for which the antifield independent
part will read
\be
  B_a(\phi, \theta)\gh C^a=B_a(\phi)\gh C^a+\gh O(\theta)=
     b_k A^k_a(\phi)\gh C^a+\gh O(\theta).
\label{extra anom}
\ee
The contribution $\gh O(\theta)$ results to be relevant
at order $\hbar^{3/2}$ or higher. Indeed, if we undo
\bref{teta trans}, $\gh O(\theta)$ become $\gh O(\sqrt\hbar\theta)$,
contributing thus at higher orders in $\hbar$.
Thus, in one loop considerations, $\gh O(\theta)$ terms can be
discarded. Quantums effects of the extra
variables are then expected to be realized as
shifts or ``renormalizations'' of the original coefficients of the
anomalies $A_a(\phi)$,
$a_k\rightarrow \tilde a_k=(a_k+b_k)$,
and, as a byproduct, of the coefficients
of all the quantities directly related with them,
like the Wess-Zumino term $M^{(0)}_1$ or the kinetic operator $D_{ab}$.

Let us now regularize the extended theory along the lines
of sect.2.3. As the analogous of the regulated generating
functional \bref{reg gen func}, we take
\be
  Z_{\rm reg}(J,K;j,\theta^*)=\int\gh D \tilde\Phi \gh D \tilde\chi
  \restric{\exp\left\{\frac{i}{\hbar}
  \left[\widehat W +J_A\Phi^A+j_a\theta^a\right] \right\}}
  {\tilde\chi^*=0},
\label{reg gen func 2}
\ee
with $\widehat W$ given by
\be
  \widehat W =\left[\tilde W_0 + W_{\rm PV}
  +\sqrt\hbar \tilde M_{1/2}+ \hbar\tilde M\right],
\label{w tilde 2}
\ee
and where $\tilde\Phi\equiv\{\tilde\Phi^\mu\}$ stands for the complete
set of fields $\{\Phi^A,\theta^a\}$ and
$\tilde\chi\equiv\{\tilde\chi^\mu\}$ for their associated
PV fields $\{\chi^A,\chi^a\}$.

According to the above discussion, in
the new proper classical action $\tilde W_0$ in \bref{w tilde 2},
the $\theta^a$ kinetic operator should be taken as
\be
   \widehat{D}_{ab}= c_k D^k_{ab},\quad\quad
   \mbox{\rm with}\quad\quad
   D^k_{ab}=
  \left(\frac{\partial A^k_a}{\partial\phi^i}R^i_b\right),
\label{d wide}
\ee
i.e., the original coefficients $a_k$ in \bref{47}
has been relaxed to $c_k$ and
should be determined in the regularization procedure.
$\tilde M_{1/2}$, on the other hand,
is expected to be related with the renormalization of the
original background term $M_{1/2}$ in \bref{wm12}, while $\tilde M$ is a
suitable counterterm taking care of possible dependences
on the auxiliar sector and on the antifields.

Finally, the PV action
$\restric{W_{\rm PV}}{\chi^*=0}$ in \bref{w tilde 2} is determined from
$\tilde W_0$ in \bref{w tilde 2} by means of \bref{orig pv action}
$$
   \restric{W_{\rm PV}}{\chi^*=0}
   =\restric{W^{(0)}_{\rm PV}}{\chi^*=0}+W_M=
    \frac12 \chi^\mu (\tilde T\tilde R)_{\mu\nu} \chi^\nu
    -\frac12 M \tilde\chi^\mu \tilde T_{\mu\nu} \tilde\chi^\nu,
$$
with $(\tilde T\tilde R)_{\mu\nu}$ defined according to \bref{s der} as
\be
  (\tilde T\tilde R)_{\mu\nu}=
  \left(\lder{}{\tilde\Phi^\mu}\rder{}{\tilde\Phi^\nu}\tilde W_0\right),
\label{new ktr}
\ee
and where we choose the mass term $W_M$ with no mixing between
the original PV fields $\chi^A$ and the extra ones $\chi^a$ and
containing the mass matrix $T_{AB}$ used in the regularization of the
original theory, i.e.,
\be
   W_M=-\frac12 M \tilde\chi^\mu \tilde T_{\mu\nu} \tilde\chi^\nu=
   -\frac12 M \left(\chi^A T_{AB} \chi^B
                  + \chi^a T_{ab} \chi^b\right).
\label{new mass}
\ee

Now, an straightforward application of the semiclassical expansion to
\bref{reg gen func 2} yields the effective action of the extended theory up
to one loop
\be
  \tilde \Gamma= \tilde W_0+\sqrt\hbar \tilde M_{1/2}+ \hbar\tilde M
  + \frac{i\hbar}2{\rm Tr}\ln\left[\frac{(\tilde T\tilde R)}
  {(\tilde T\tilde R)-\tilde T M}\right],
\label{ext gamma 1}
\ee
while from its corresponding BRST Ward identity \bref{wi}
the following anomaly arises
\be
  -i \hbar\tilde{\gh A}=
  \sqrt\hbar(\tilde M_{1/2}, \tilde W_0)
  -i\hbar\left[(\Delta \tilde W_0)_{\rm reg}
   +i(\tilde M, \tilde W_0)+
   \frac{i}{2}(\tilde M_{1/2}, \tilde M_{1/2})\right].
\label{reg anomaly 2}
\ee
Finally,
the regularized expression of $(\Delta \tilde W_0)$ (or, equivalently, of
$(\Delta \hat S_{\rm ext})$ since both share the same unregularized form)
turns out to be, once again, the BRST variation of the trace part in
\bref{ext gamma 1}
\be
  (\Delta \tilde W_0)_{\rm reg}=
  \tilde\delta\left\{-
  \frac{1}{2}{\rm Tr}\ln\left[\frac{\tilde R}{\tilde R-M}\right]\right\}
  = {\rm Tr}\left[-\frac12(\tilde R^{-1}\tilde \delta \tilde R)
   \frac{1}{(1-\tilde R/M)}\right],
\label{a 22}
\ee
where $\tilde\delta$ is now the BRST transformation in the extended space
generated by $\tilde W_0$, $\tilde\delta F= (F,\tilde W_0)$.

Let us now investigate the form of
$\tilde R$ along the lines of sect.3 to gain insight into the
structure of \bref{a 22}.
By direct inspection of $\tilde W_0$ in \bref{w tilde 2},
$(\tilde T\tilde R)_{\mu\nu}$ \bref{new ktr} is seen to be
$$
  (\tilde T\tilde R)_{\mu\nu}=
   \left(\begin{array}{cc}
    (TR)_{AB}(\Phi,K)&\mbox{}\\
   \mbox{}& -i \widehat{D}_{ab}(\phi)
   \end{array}\right)
   +\gh O(\theta)+\gh O(\theta^*),
$$
with $(TR)_{AB}$ the original massless kinetic operator
\bref{orig pv kin}. Then, the inverse of the new mass matrix
in \bref{new mass} leads to the extended regulator $\tilde R$
$$
  (\tilde T^{-1})^{\mu\rho}(\tilde T\tilde R)_{\rho\nu}
  \equiv\tilde R^\mu_{\;\nu}=
   \left(\begin{array}{cc}
    R^A_{\,B}&\mbox{}\\
   \mbox{}& -i(T^{-1})^{ac} \widehat{D}_{cb}
   \end{array}\right)
   +\gh O(\theta) +\gh O(\theta^*).
$$
Using now the expansion \bref{exp r min} for
the original regulator $R$ and assuming an
expansion for $T_{ab}$ of the form
$T_{ab}= T_{0,ab}(\phi)+\ldots$, with $T_{0,ab}(\phi)$ invertible, we
obtain a similar expansion for $\tilde R$
\be
  \tilde R=[\tilde{\gh R}(\phi)+\gh O(\theta)]+\widehat{\tilde R},
  \quad\quad\mbox{with}\quad\quad
  \tilde{\gh R}^\mu_{\;\nu}(\phi)=
   \left(\begin{array}{cc}
    \gh R^A_{\;B}(\phi)&\mbox{}\\
   \mbox{}& -i(T^{-1}_0)^{ac} \widehat{D}_{cb}(\phi)
   \end{array}\right),
\label{gh R 2}
\ee
and where $[\tilde{\gh R}(\phi)+\gh O(\theta)]$ plays now an analogous role
as that of $\gh R(\phi)$ \bref{exp r min} in the original theory.
Similar expansions are shared by
$\tilde R^{-1}$ and $\tilde\delta\tilde R$ due to the
linearity in $\theta^a$ of $\tilde\delta\theta^a$.
In the end, plugging all these results in
$(\Delta \tilde W_0)_{\rm reg}$ \bref{a 22},
the following expression is obtained
\be
  (\Delta \tilde W_0)_{\rm reg}= \left\{
   {\rm Tr}\left[-\frac12(\tilde{\gh R}^{-1}\tilde\delta\tilde{\gh R})
   \frac{1}{(1-\tilde{\gh R}/M)}\right] + \gh O(\theta)\right\}+
   (\widehat{\Delta \tilde W_0})_{\rm reg}+(\gh M',\tilde W_0),
\label{a 23}
\ee
with $(\widehat{\Delta \tilde W_0})_{\rm reg}=\gh O(K,\theta^*)$ and
where each one of the terms in \bref{a 23} is
$\tilde\delta$ invariant by itself.

Let us restrict now to the study of the antifield independent part in
\bref{a 23}. $\gh O(\theta)$,
as argued below, can simply be discarded. On the
other hand, the diagonal
structure of $\tilde{\gh R}$ \bref{gh R 2} yields the following
decomposition for the trace in eq.\bref{a 23}
\be
  {\rm Tr}\left[-\frac12(\tilde{\gh R}^{-1}\tilde\delta\tilde{\gh R})
   \frac{1}{(1-\tilde{\gh R}/M)}\right]=
  {\rm Tr}\left[-\frac12(\gh R^{-1}\delta\gh R)
   \frac{1}{(1-\gh R/M)}\right]+
  {\rm Tr}\left[-\frac12(\gh R_\theta^{-1}\delta\gh R_\theta)
   \frac{1}{(1-\gh R_\theta/M)}\right],
\label{new dec}
\ee
with $\gh R_\theta$ defined as
\be
   (\gh R_\theta)^a_{\;b}(\phi)=-i
   (T^{-1}_{0})^{ac} \widehat{D}_{cb}(\phi),
\label{r teta}
\ee
and where now distinction between $\delta$ or $\tilde\delta$
is irrelevant as they share the same form for the
classical fields:
$ \delta\phi^i= \tilde\delta\phi^i= R^i_\alpha\gh C^\alpha$.

The first trace in the right-hand side of \bref{new dec} is
the original anomaly, $A_a(\phi)\gh C^a$, whereas
the second trace should be considered as the
antifield independent contribution \bref{extra anom} to
the anomaly coming from the extra fields. This
second term could produce type $A$ anomalies unless an $A$ invariant
regulator $\gh R_\theta$ \bref{r teta} is found%
\footnote{Invariance of regulators should be understood up to terms of the
form $[\gh R, G]$, i.e., they are invariant as far as they yield vanishing
anomalies.}.
In appendix C we argue that the transformation property
of $T_{0,ab}(\phi)$ under the $A$ subgroup
$$
   \delta_{(A)} T_{0,ab}=
   \left(T_{0,ac} T^c_{bA}+ T_{0,bc} T^c_{aA}\right)\veps^A,
$$
is a sufficient condition to get this result.
Assuming then that such a mass matrix has been found, the
final form for $(\Delta \tilde W_0)_{\rm reg}$ \bref{a 23}
reads
$$
   (\Delta \tilde W_0)_{\rm reg}=
   (A_a+B_a)(\phi)\gh C^a +
   (\widehat{\Delta \tilde W_0})_{\rm reg}+(\gh M',\tilde W_0).
$$

It is straightforward now to determine
$\tilde M_{1/2}$ and $\tilde W_0$ which yield the
vanishing of the antifield independent part
of $\tilde{\gh A}$ \bref{reg anomaly 2}.
Indeed, the $\hbar$ part of \bref{reg anomaly 2} vanishes for
\be
    \tilde M_{1/2}=
    -\left[ \theta^*_a\gh C^a+i (A_a+B_a)(\phi)\theta^a\right]=
    -\left[ \theta^*_a\gh C^a+i \tilde A_a(\phi)\theta^a\right].
\label{new back}
\ee
with $\tilde A_a=(a_k+b_k)A^k_a$,
whereas vanishing of the $\sqrt\hbar$ term
is acquired for $\widehat{D}_{ab}$ \bref{d wide} of the form
\be
  \widehat{D}_{ab}=
  \left(\frac{\partial\tilde A_a}{\partial\phi^i}R^i_b\right)=
  \tilde D_{ab}.
\label{d k 2}
\ee

Equations \bref{new back} and \bref{d k 2} express hence the
conditions for the vanishing of the antifield independent part of
$\tilde \gh A$ and, as a
consequence, for the (partial) fulfillment of
\bref{wi} for $\tilde \Gamma$ \bref{ext gamma 1}
up to one loop in the antifield independent sector.
Implementation of these
conditions leads to the one--loop renormalized action
$$
  \tilde W_0+\sqrt\hbar\tilde M_{1/2} +\hbar \tilde M=
   \left[\hat S(\Phi, K) -\frac{i}2 \theta^a \tilde D_{ab}\theta^b
  +\theta^*_aT^a_{Bb}\theta^b\gh C^B \right]
  -\sqrt\hbar\left[\theta^*_a\gh C^a+i \tilde A_a\theta^a\right]
  +\hbar \tilde M,
$$
which, in terms of the original extra variables \bref{teta trans} becomes
the solution \bref{tilde w} of the regularized quantum master equation at
one-loop level in the extended formalism, i.e.,
\bea
   &\displaystyle{\left[\hat S(\Phi, K) -\theta^*_a\left(\gh C^a-
   T^a_{Bb}\theta^b\gh C^B\right) \right]
   -i\hbar\left[\tilde A_a\theta^a+
   \frac{1}2 \theta^a \tilde D_{ab}\theta^b\right]
  +\hbar \tilde M=}&
\nonumber\\
  &\displaystyle{\hat S_{\rm ext} +\hbar[\tilde M^{(0)}_1 +\tilde M]=
           \hat S_{\rm ext} +\hbar\tilde M_1\equiv\tilde W.}&
\label{late w}
\eea

In summary, from \bref{late w} it is concluded that, at source independent
level, the effect of the extra degrees of freedom is realized as a shift or
renormalization of the coefficients $a_k$ of the original Wess-Zumino term
to $(a_k+b_k)$,
as argued before. This ends the description of the regularization procedure
in the extended theory.

\section{Invariant Pauli-Villars regularization in the extended
         configuration space}

\hspace{\parindent}%
Anomalous gauge theories are known to suffer
from the absence of BRST (or gauge) invariant regulators in the original
configuration space. Within the above extended formalism, instead,
they give rise to BRST invariant theories up to one loop in the antifield
independent sector.
This fact suggests the existence of PV
invariant regulators $\tilde{\gh R'}(\phi,\theta)$ in
the extended formalism.
Such possibility has been considered in \cite{hid}%
\footnote{Invariant PV regularizations of this type
were earlier considered, for chiral gauge theories, in \cite{Bal87}.
More recently, a similar invariant PV regularization has also been
used in \cite{FS91} in the quantization of the two dimensional
chiral Higgs model.},
although this formulation differs in spirit from ours.
In this section we shall show, first, how to construct such invariant
PV regulators and second,
how a natural interpretation arises of the
Wess-Zumino action as the local counterterm
interpolating between invariant and noninvariant regularizations.

\subsection{Completely anomalous gauge theory}

\hspace{\parindent}%
To illustrate the construction,
let us consider first of all the case $r=m$ in \bref{rank a}.
In the extended theory
the combinations $F^i(\phi,\theta)$ result to be gauge (or BRST) invariant.
Therefore, an invariant regulator
${\gh R}'(\phi,\theta)$ can be built up from a non-invariant one
${\gh R}(\phi)$ by the simple rule of substituting the fields $\phi^i$
in ${\gh R}$ by their gauge transformed $F^i(\phi,\theta)$, that is,
\be
   {\gh R}'(\phi,\theta)\equiv
   {\gh R}(F(\phi,\theta))\Rightarrow
   \delta{\gh R}'=0.
\label{orig inv r}
\ee

The construction of invariant regulators in this way turns out to be
a useful tool to ``integrate'' anomalies and obtain
the Wess-Zumino action. This observation is based
on the following facts. First of all, eq.\bref{eqwz} for
$M^{(0)}_1(\phi,\theta)$ can be interpreted
as the expression relating the anomalies
$$
i{\gh B}_1(\phi,\gh C^\alpha)=0, \quad\quad
i{\gh B}_0(\phi,\gh C^\alpha)=
i  A_\alpha(\phi)\gh C^\alpha,
$$
arising in the invariant $(1)$ and non-invariant $(0)$
regularizations, through the BRST variation of a local counterterm in the
extended configuration space, i.e.,
$$
  i\gh B_1- i\gh B_0=
  -i A_\alpha(\phi)\gh C^\alpha=\delta(-M^{(0)}_1(\phi,\theta))
   \Longleftrightarrow (M^{(0)}_1, \hat S_{\rm ext})=
   iA_\alpha(\phi)\gh C^\alpha.
$$
On the other hand, these two regularizations are connected by
the interpolation
\be
  {\gh R}(t)={\gh R}(F(\phi,\theta t)),\quad\quad t\in [0,1].
\label{inter 1}
\ee
Under such conditions, we can apply the results in appendix B
and take for $M^{(0)}_1(\phi,\theta)$
expression \bref{m zero r} adapted to this case, namely
\be
   M^{(0)}_1(\phi,\theta)=-i\int^1_0\dif t\, {\rm Tr}
   \left\{-\frac12 \left[{\gh R}^{-1}(F(\phi,\theta t))\partial_t
   {\gh R}(F(\phi,\theta t))\right]\frac{1}
   {\left(1-{\gh R}(F(\phi,\theta t))/M\right)}\right\}.
\label{new m1}
\ee

Now, explicit computation of the
$\partial_t$ derivative of ${\gh R}(t)$ \bref{inter 1}
yields, after use of the Lie equation for $F^i(\phi,\theta)$ \cite{GP93}
$$
     \partial_t {\gh R}(F(\phi,\theta t))=
     \left(\der{{\gh R}}{\phi^i}\right)(F)
     \partial_t F^i(\phi,\theta t)=
     \left(\der{{\gh R}}{\phi^i} R^i_\beta\right)(F)
     \lambda^\beta_\alpha(\theta t,\phi)\theta^\alpha=
     (\delta_\beta {\gh R})(F)
     \lambda^\beta_\alpha(\theta t,\phi)\theta^\alpha,
$$
where $(\delta_\beta {\gh R})$ stands for the BRST (or gauge)
variation of ${\gh R}$ having dropped out the ghosts $\gh C^\beta$
(or the gauge parameters
$\veps^\beta$). Upon substitution of this result in \bref{new m1}
\be
   M^{(0)}_1(\phi,\theta)=-i\int^1_0\dif t\, {\rm Tr}
   \left[-\frac12 ({\gh R}^{-1}\delta_\beta {\gh R})\frac{1}
   {\left(1-{\gh R}/M\right)}\right](F)
     \lambda^\beta_\alpha(\theta t,\phi)\theta^\alpha,
\label{m1int}
\ee
we recognize in the trace factor the form \bref{a alpha} for the
anomaly with argument $F^i(\phi,\theta t)$, so that \bref{m1int}
acquires the form \bref{wz} of the Wess-Zumino term for the original
theory. This expression was previously derived in \cite{GP93}
using a different approach.

Therefore, from this construction
a new interpretation \cite{hid} of the Wess-Zumino term arise:
it is the local counterterm giving the interplay
between the original, non-invariant regularization and the new invariant
one \bref{orig inv r}.

\subsection{Anomalous free gauge subgroup}

\hspace{\parindent}%
The situation just considered is very restrictive since
certain theories possess regulators preserving a subgroup (the $A$ part)
of the gauge transformations.
Therefore, a modification of the above proposal should be considered.
We restrict our analysis to the case in which the anomalous ($a$) sector
is a subgroup.

Assume then that the original regulator $\gh R(\phi)$ satisfies
\bref{deltas r 1}. The analogous of the invariant objects
$F^i(\phi,\theta)$ considered
in the previous case are now the combinations
$\tilde F^i(\phi,\theta^a)=F^i(\phi,\theta^a,\theta^A=0)$,
with transformation laws
$$
  \delta_{(a)}\tilde F^i=0,\quad\quad
  \delta_{(A)}\tilde F^i= R^i_B(\tilde F)M^B_A\veps^A,
$$
and with $M^B_A$ an invertible matrix whose form is irrelevant for our
purposes \cite{GP93}.
Then, the desired invariant regulator $\gh R'(\phi,\theta)$ turns out to be
in terms of the original one
\be
   \gh R'(\phi,\theta^a)=\gh R(\tilde F(\phi,\theta)).
\label{inv r 2}
\ee

Indeed, invariance of $\gh R'(\phi,\theta)$
under $a$ transformations comes from the $a$ invariance of $\tilde F^i$,
$\delta_{(a)}\tilde F^i=0$, as in the previous case, while for the $A$ part
it is
$$
  \delta_{(A)}\gh R'=
  \left(\derpar{\gh R}{\phi^i} R^i_A\right)(\tilde F)\veps^A=
  (\delta_{(A)}\gh R)(\tilde F)\veps^A=
   \left\{\begin{array}{l}
    0  \\
    {[\gh R, G_B]}(\tilde F)M^B_A\veps^A
   \end{array}\right.
$$
the result being now a direct consequence of the invariance
\bref{deltas r 1} of $\gh R$ under the $A$ subgroup.

Finally, the Wess-Zumino term $M^{(0)}_1(\phi,\theta^a)$, interpreted again
as the counterterm relating the anomalies
$$
i{\gh B}_1(\phi,\gh C^a)=0, \quad\quad
i{\gh B}_0(\phi,\gh C^a) = i A_a(\phi)\gh C^a,
$$
obtained using invariant $(1)$ and non-invariant $(0)$ regulators, can be
constructed along the previous lines simply by substituting the above
quantitites by those corresponding to the $a$ subgroup
as a group by itself. In particular, the
interpolating regularization between $\gh R(\phi)$ and
$\gh R'(\phi,\theta^a)$ reads now
\be
  \gh R(t)=\gh R(\tilde F(\phi,\theta^a t)),\quad\quad t\in [0,1],
\label{inter 2}
\ee
yielding at the end the same form for
$M^{(0)}_1(\phi,\theta^a)$ worked out in \cite{GP93}
for this particular case.

\section{Example: the Bosonic String}

\hspace{\parindent}%
In this section we illustrate the use of the extended formalism
by applying it to the bosonic string%
\footnote{Quantization of the bosonic string as an anomalous
gauge theory along the lines of the hamiltonian BRST formalism \cite{BFV}
has recently been considered in \cite {Fujiwara}.}.
In this way we will see
that a natural interpretation arises of the well-known shift
of the numerical coefficient $(26-D)$ in front of the Liouville action to
$(25-D)$, in agreement with \cite{DDK}%
\footnote{For earlier comments about this shift, see ref.\cite{Mar83}.}.
This model will also serve to exemplify
the method proposed in sect.5 for constructing invariant regulators and
Wess-Zumino actions in the extended configuration space.

\subsection{Regularization of the original theory}

\hspace{\parindent}%
The bosonic string is an example of a gauge theory in which part of the
gauge group can be kept anomaly free while the anomalous part can be chosen
to be an abelian subgroup. The classical action for this system
$$
   S_0=\int\dif^2\xi\left[-\frac12\sqrt{g}g^{\alpha\beta}
       \partial_\alpha X\partial_\beta X\right],\quad\quad
        \mbox{with}\quad
  g\equiv-\det g_{\alpha\beta},
$$
describes $D$ bosons $X^\mu(\xi)$ coupled to
the gravitational field $g_{\alpha\beta}$ in two dimensions and posseses
the following (infinitesimal) gauge transformations
\bea
     \delta X^\mu&=&v^\alpha\partial_\alpha X^\mu,
\nonumber\\
     \delta g_{\alpha\beta}&=&\nabla_\alpha v_\beta+\nabla_\beta
     v_\alpha+\lambda g_{\alpha\beta},
\nonumber
\eea
which split into two subgroups: Weyl transformations $(\lambda)$ and
diffeomorphisms $(v^\alpha)$.

Direct application of the FA formalism yields as proper
solution of \bref{cme}
\be
    S= S_0+\int\dif^2 \xi
    \left[ X^*\gh C^\alpha\partial_\alpha X+
    g^{*\alpha\beta}\left(
    \nabla_\alpha \gh C_\beta+\nabla_\beta \gh C_\alpha+\gh C
     g_{\alpha\beta}\right)
    -\gh C^*_\beta\gh C^\alpha\partial_\alpha\gh C^\beta
    -\gh C^* \gh C^\alpha\partial_\alpha\gh C+
    b^*_{\alpha\beta} d^{\alpha\beta} \right],
\label{prop bos}
\ee
where $\{\gh C^\alpha, \gh C\}$ are the
diffeomorphisms and Weyl ghosts, respectively,
$\{X^*,g^{*\alpha\beta},\gh C^*_\beta, \gh C^*\}$ the
antifields of the minimal sector fields and
$\{b^{\alpha\beta}, d^{\alpha\beta}; b^*_{\alpha\beta},
d^*_{\alpha\beta}\}$ the fields and antifields of the auxiliar sector.

A usual gauge-fixing fermion is
$\Psi=-1/2\,b^{\alpha\beta} (g_{\alpha\beta}-h_{\alpha\beta})$, with
$h_{\alpha\beta}$ a given background metric.
Using now \bref{gf}, the antifields acquiring a shift are
$$
  g^{*\alpha\beta}\rightarrow
  g^{*\alpha\beta} -\frac12 b^{\alpha\beta},\quad\quad
  b^*_{\alpha\beta}\rightarrow
  b^*_{\alpha\beta}-\frac12 (g_{\alpha\beta}-h_{\alpha\beta}),
$$
so that in the new gauge-fixed basis, $S$ \bref{prop bos} adopts the form
\bea
    &&\hat S(\Phi,K)= \int\dif^2 \xi\left\{\left[\frac12 X\Box X
    -\frac12 b^{\alpha\beta}
    \left(\nabla_\alpha \gh C_\beta+\nabla_\beta \gh C_\alpha+\gh C
     g_{\alpha\beta}\right)
     -\frac12 d^{\alpha\beta}
    (g_{\alpha\beta}-h_{\alpha\beta})\right]+\right.
\nonumber\\
    &&\left.\hspace{22mm}\left[
    X^*\gh C^\alpha\partial_\alpha X+
    g^{*\alpha\beta}\left(
    \nabla_\alpha \gh C_\beta+\nabla_\beta \gh C_\alpha+\gh C
     g_{\alpha\beta}\right)
   -\gh C^*_\beta\gh C^\alpha\partial_\alpha\gh C^\beta
    -\gh C^* \gh C^\alpha\partial_\alpha\gh C+
    b^*_{\alpha\beta} d^{\alpha\beta} \right]\right\}
\nonumber\\
    &&\hspace{15mm}=S_\Sigma(\Phi)+K_A R^A,
\label{sgf bos}
\eea
with $R^A$ the BRST transformation of the field $\Phi^A$ and
where, for simplicity, antifields $\Phi^*_A$
and BRST sources $K_A$ are identified. Also, the kinetic operator for the
matter fields $X^\mu$ in \bref{sgf bos} is defined by
$ \Box= \partial_\alpha(\sqrt g g^{\alpha\beta}\partial_\beta)
 =\sqrt g g^{\alpha\beta}\nabla_\alpha\nabla_\beta$.

The form of the gauge-fixed action $S_\Sigma$ in \bref{sgf bos}
suggests some field redefinitions in order to distinguish propagating and
non-propagating fields. Indeed, by introducing a new symmetric,
traceless field $\tilde b^{\alpha\beta}$ and a new pair of ghosts
$\tilde b$, $\tilde{\gh C}$, related to the old ones
$b^{\alpha\beta}$, ${\gh C}$ by
$$
  b^{\alpha\beta}=\tilde b^{\alpha\beta}+\frac12 g^{\alpha\beta} \tilde b,
  \quad\quad
  \gh C=\tilde{\gh C}-\nabla_\alpha\gh C^\alpha,
$$
the gauge-fixed action adopts the form
$$
  S_\Sigma(\Phi)= \int\dif^2 \xi \left[\frac12 X\Box X
    -\frac12 \tilde b^{\alpha\beta}
    \left(\nabla_\alpha \gh C_\beta+\nabla_\beta \gh C_\alpha\right)
    -\frac12 \tilde b\tilde{\gh C}
     -\frac12 d^{\alpha\beta}
    (g_{\alpha\beta}-h_{\alpha\beta})\right],
$$
and allows to identify the ghosts $\tilde b$, $\tilde{\gh C}$ and the
fields $d^{\alpha\beta}$, $g_{\alpha\beta}$ as non-propagating.
These fields will not occur in loops and their contribution to the
anomaly is expected to vanish.

Now, let us pass to analyze the regularized expression of
$\Delta \hat S$. The regulator $R(\Phi, K)$ in \bref{a 2 quad}
is determined from the PV massless
kinetic operator $(TR)_{AB}$ \bref{orig pv kin}
and the PV mass matrix $T_{AB}$. In the basis
$\{X,\tilde b^{\alpha\beta},\gh C^\alpha;
d^{\alpha\beta},g_{\alpha\beta},\tilde b,\tilde{\gh C}\}
\equiv\{{\rm p; np}\}$, $(TR)_{AB}$ adopts the form
$$
  (TR)_{AB}(\Phi,K)=
   \left(\begin{array}{cc}
    (TR)_{\rm p}(g_{\alpha\beta})&\mbox{}\\
   \mbox{}& (TR)_{\rm np}
   \end{array}\right) +\gh O(X)
   +\widehat{(TR)}_{AB}(\Phi, K),
$$
where the corresponding {\it invertible} blocks for propagating and
non-propagating fields read
$$
  (TR)_{\rm p}(g_{\alpha\beta})=
   \left(\begin{array}{ccc}
    \Box &\mbox{}&\mbox{}\\
   \mbox{}& \mbox{}& (TR)_{\alpha\beta\gamma}\\
   \mbox{}& -(TR)_{\alpha\beta\gamma}&\mbox{}
   \end{array}\right),
   \quad\quad
  (TR)_{\rm np}=
   \left(\begin{array}{cccc}
    0&1&\mbox{}&\mbox{}\\
    1&0&\mbox{}&\mbox{}\\
    \mbox{}&\mbox{}&0&-1/2\\
    \mbox{}&\mbox{}&1/2&0
   \end{array}\right).
$$
For the mass matrix $T_{AB}$ one can take
$$
  T_{AB}=
   \left(\begin{array}{cc}
    T_{\rm p}(g_{\alpha\beta})&\mbox{}\\
   \mbox{}& T_{\rm np}
   \end{array}\right),
\quad\quad\mbox{with}\quad
  T_{\rm p}=
   \left(\begin{array}{cc}
    \sqrt g&\mbox{}\\
   \mbox{}& T_{\rm gh}(g_{\alpha\beta})
   \end{array}\right),
$$
and where $T_{\rm np}$ is chosen to be a constant, invertible matrix, its
form being irrelevant. In the end,
$R(\Phi, K)$ is seen to have an expansion of the form
\bref{exp r min}, as expected
\be
   R^A_{\,B}(\Phi,K)=\left[
   \left(\begin{array}{cc}
    \gh R_{\rm p}(g_{\alpha\beta})&\mbox{}\\
   \mbox{}& \gh R_{\rm np}
   \end{array}\right)+ \gh O(X)\right]
   +\widehat R^A_{\,B}(\Phi,K),
\quad\mbox{with}\quad
  \gh R_{\rm p}=
   \left(\begin{array}{cc}
    \frac{1}{\sqrt g}\Box&\mbox{}\\
   \mbox{}& \gh R_{\rm gh}(g_{\alpha\beta})
   \end{array}\right),
\label{reg bos 1}
\ee
and where the regulator $\gh R_{\rm np}$ for the non-propagating fields
results to be a {\it constant}, invertible matrix. The explicit
forms of $(TR)_{\alpha\beta\gamma}$, $T_{\rm gh}$ and $\gh R_{\rm gh}$
are not necessary for our immediate purposes and can
be found in \cite{DTNP88}, from where some results will be borrowed.

Now, let us restrict to the
analysis of the antifield independent part of
$(\Delta \hat S)_{\rm reg}$, \bref{a alpha}.
The splitting of $\gh R(\phi)$ in \bref{reg bos 1} in an
invertible part containing only $g_{\alpha\beta}$ plus
terms $\gh O(X)$, together with the property
$\delta\gh O(X)=\gh O(X)$, leads to the splitting of
\bref{a alpha} as
\be
   {\rm Tr}\left[-\frac12(\gh R^{-1}\delta\gh R)
   \frac{1}{(1-\gh R/M)}\right]+\gh O(X),
   \quad\mbox{where}\quad
   \gh R(g_{\alpha\beta})=
   \left(\begin{array}{cc}
    \gh R_{\rm p}(g_{\alpha\beta})&\mbox{}\\
   \mbox{}& \gh R_{\rm np}
   \end{array}\right),
\label{delta bos}
\ee
each one of the terms being separately BRST invariant.
Well-known cohomological results ensure that BRST invariant $\gh O(X)$
terms are BRST trivial in local cohomology, so that
$\gh O(X)$ terms in \bref{delta bos} can eventually be written as the
BRST variation of a suitable counterterm.

We can thus restrict the analysis of \bref{delta bos} to the contribution
coming from $\gh R(g_{\alpha\beta})$.
First of all, the constant block $\gh R_{\rm np}$
in \bref{delta bos} produces, upon
the $\delta$ variation of $\gh R$, a vanishing entry for
this part. The relevant part in \bref{delta bos} then becomes
\be
   {\rm Tr}\left[-\frac12(\gh R_{\rm p}^{-1}\delta\gh R_{\rm p})
   \frac{1}{(1-\gh R_{\rm p}/M)}\right],
\label{delta bos 2}
\ee
yielding a vanishing contribution of the non-propagating
fields to the anomaly. On the other hand,
the diagonal block structure of
$\gh R_{\rm p}$ \bref{reg bos 1}
splits \bref{delta bos 2} into two separate contributions
\be
   {\rm Tr}\left[-\frac12(\gh R_{\rm m}^{-1}\delta\gh R_{\rm m})
   \frac{1}{(1-\gh R_{\rm m}/M)}\right]+
   {\rm Tr}\left[-\frac12(\gh R_{\rm gh}^{-1}\delta\gh R_{\rm gh})
   \frac{1}{(1-\gh R_{\rm gh}/M)}\right],
\label{split2}
\ee
coming each one from the matter field sector and from the ghost fields
$(\tilde b^{\alpha\beta} ,\gh C^\alpha)$ sector, respectively,
and with the matter regulator $\gh R_{\rm m}$ given by
\be
  \gh R_{\rm m}=\frac{1}{\sqrt g} \Box=
 \frac{1}{\sqrt g}\partial_\alpha(\sqrt g g^{\alpha\beta}\partial_\beta)
 =g^{\alpha\beta}\nabla_\alpha\nabla_\beta.
\label{reg corda}
\ee

Now, for definiteness, let us analyze the matter contribution
in \bref{split2} as an example of the use of the form \bref{a 2 quad}
for $(\Delta \hat S)_{\rm reg}$. First, the BRST variation of
$\gh R_{\rm m}$ \bref{reg corda} reads
$$
  \delta \gh R_{\rm m}=-\gh R_{\rm m} \gh C-[\gh R_{\rm m}, G],\quad\quad
  G\equiv \gh C^\alpha\partial_\alpha,
$$
so that, whereas it transforms in the ``appropriate'' way under
diffeomorphisms, its Weyl variation neither is zero nor it can be written
in commutator form. The whole gauge group splits thus into
two subgroups, one of them, diffeomorphisms, anomaly
free and the other, the abelian Weyl subgroup, anomalous.
Then, upon substitution of this result in the first term of \bref{split2},
we obtain
\be
   {\rm Tr}\left[-\frac12(\gh R_{\rm m}^{-1}\delta\gh R_{\rm m})
   \frac{1}{(1-\gh R_{\rm m}/M)}\right]=
  {\rm Tr}\left[\frac12 \gh C
   \left(1-\frac{\Box}{\sqrt g M}\right)^{-1}\right]\sim
  {\rm Tr}\left[\frac12 \gh C\,
   \exp\left\{\frac{\Box}{\sqrt g M}\right\}\right],
\label{w1}
\ee
which turns out to be, using well-known results on the
calculation of the Weyl anomaly%
\footnote{Traces over continuous indices involved in expressions like
          \bref{w1} should be computed in the euclidian space \cite{VP93}.
          Coming back to the Minkowsky space causes the appearence of $-i$
          factors, as the one in front of \bref{mat anom}.}
\cite{Fuji}
\be
  -i\int\dif^2\xi
   \left[\left(\frac{D}{8\pi}\right) M\sqrt{g}-
   \left(\frac{D}{48\pi}\right)\sqrt g R\right]\gh C\equiv
   A_{\rm m}(g_{\alpha\beta})\cdot\gh C,
\label{mat anom}
\ee
where $R$ is the scalar curvature and $A_{\rm m}(g_{\alpha\beta})$ stands
for the contribution of the matter sector to the Weyl anomaly.
Contributions to
\bref{delta bos 2} coming from the ghost sector can be treated in
the same way using the regulator $\gh R_{\rm gh}$ obtained from
$(TR)_{\alpha\beta\gamma}$ and $T_{\rm gh}$ proposed in
\cite{DTNP88}. The net effect of these contributions is a shift in
the above numerical coefficients by
$$
   \frac{-D}{48\pi}\rightarrow \frac{26-D}{48\pi},\quad\quad
   \frac{D}{8\pi}\rightarrow \frac{D-2}{8\pi}.
$$

Finally,
the divergent pieces arising after regularization in \bref{split2}
(partly displayed in \bref{mat anom}) are
seen to be absorbed by the BRST variation of a suitable local counterterm
\be
  \int\dif^2\xi
   \left[\left(\frac{D-2}{8\pi}\right) M\sqrt{g}\,\gh C\right]=
   \delta\left\{ \int\dif^2\xi
   \left[\left(\frac{D-2}{8\pi}\right) M\sqrt{g}\right]\right\},
\label{div count}
\ee
so that in the end, the antifield independent part of the original anomaly
to deal with becomes
\be
  \int\dif^2\xi
   \left[\left(\frac{26-D}{48\pi}\right)\sqrt g R\right]\gh C\equiv
   i A(g_{\alpha\beta})\cdot\gh C.
\label{weyl anom}
\ee

\subsection{Extended theory, invariant regulator and Wess-Zumino term}

\hspace{\parindent}
Once the anomalous character of the theory has been verified,
the construction of the extended field-antifield formalism
goes as follows \cite{GP93}. Since only the one parametric Weyl group is
anomalous, no rank troubles arise and we are led to introduce a new scalar
field $\theta$. Its
transformation under the action of the whole gauge group reads \cite{GP93}
\be
   \delta\theta=v^\alpha\partial_\alpha \theta-\lambda,
\label{teta bos}
\ee
yielding thus the following form of the extended action $S_{\rm ext}$
\be
  S_{\rm ext}=S+\int\dif^2\xi\,[\theta^*
   (\gh C^\alpha\partial_\alpha \theta-\gh C)].
\label{tilde s bos}
\ee

The Wess-Zumino term which corresponds to
the original Weyl anomaly \bref{weyl anom}
can now be evaluated
either by integrating the BRST
variation of the Wess-Zumino term giving \bref{weyl anom},
or by interpreting it as the counterterm (modulo
divergent pieces) relating
invariant and non-invariant regularizations. The first possibility was
contemplated in \cite{GP93}. As for the
second, we exemplify it by considering only the matter sector.
Obviously, since the ghost contribution is exactly the same up to numerical
coefficients, the Wess-Zumino term coming from this sector should also
share the same functional form.

Let us first construct an invariant regulator $\gh R'_{\rm m}$ for the
matter sector. The rule \bref{inv r 2}
yields
$$
  \gh R_{\rm m}=\frac{1}{\sqrt g} \Box\rightarrow
  \gh R'_{\rm m}=\frac{1}{(\sqrt g)'}\Box'=\frac{e^{-\theta}}{\sqrt g}
  \Box=e^{-\theta}\gh R_{\rm m}.
$$
This invariant and the non-invariant regulator
\bref{reg corda} are related by the interpolation \bref{inter 2}
\be
  \gh R_{\rm m}(t)= e^{-\theta t}\gh R_{\rm m}
  =\frac{e^{-\theta t}}{\sqrt g}\Box, \quad\quad t\in[0,1].
\label{bos int}
\ee

Now, expression
\bref{new m1} of the searched-for counterterm gives in this case
\be
   M'^{(0)}_1(g_{\alpha\beta}, \theta)=-i\int^1_0\dif t\,{\rm Tr}
   \left[\frac12\theta
   \left(1-e^{-\theta t}\gh R_{\rm m}/ M\right)^{-1}\right].
\label{mw1}
\ee
The integrand in \bref{mw1} is expression
\bref{w1} evaluated with the Weyl transformed of the original
regulator \bref{reg corda}. Hence, we have
$$
   {\rm Tr}
   \left[\frac12\theta
   \left(1-e^{-\theta t} \gh R_{\rm m}/ M\right)^{-1}\right]\sim
   {\rm Tr}\left[\frac12 \theta\,
   \exp\left\{\gh R'_{\rm m}/ M \right\}\right]=
   A_{\rm m}(g'_{\alpha\beta}(t))\cdot\theta,
$$
with $A_{\rm m}(g_{\alpha\beta})$ given by
\bref{mat anom} and where $\gh R'_{\rm m}$,
$g'_{\alpha\beta}(t)$ stands for the (finite) Weyl transformed of the
regulator $\gh R_{\rm m}$ and of the metric field with parameter
$\theta t$, i.e., eq.\bref{bos int} and
$g'_{\alpha\beta}(t)= e^{\theta t}g_{\alpha\beta}$.
Substituting all these expressions in \bref{mw1} and
performing the integration over $t$, we get
\cite{GP93}
$$
   M'^{(0)}_1(g_{\alpha\beta},\theta)=-\int\dif^2\xi
   \left\{\left[\left(\frac{-D}{48\pi}\right)
\left(\frac12\theta\Box\theta+\sqrt{g}R\theta\right)\right]
-\left[\left(\frac{D}{8\pi}\right) M\sqrt{g}\right]
+\left[\left(\frac{D}{8\pi}\right) M\sqrt{g}e^\theta\right]\right\}.
$$

Each one of the three pieces above deserves
a different interpretation. The first one is really the
contribution of the matter sector to the Wess-Zumino term.
Substitution of the coefficient $(-D)$ by $(26-D)$ in it yields thus the
complete Wess-Zumino action. The second is the part of the local
counterterm \bref{div count} whose BRST variation gives the divergent
term in \bref{mat anom}.
Finally, the third term is a BRST (or
gauge) invariant counterterm playing no role, so that
it can be simply dropped out.
In summary, the original Wess-Zumino term reads
\be
   M^{(0)}_1(g_{\alpha\beta},\theta)=\int\dif^2\xi
   \left\{\left(\frac{D-26}{48\pi}\right)
\left[\frac12\theta\Box\theta+\sqrt{g}R\theta\right]\right\},
\label{wz bos}
\ee
and it can be interpreted as the tree level Liouville action for the
bosonic string, $\theta$ being thus the Liouville field.

To conclude, a direct inspection to the form
of the $\theta$ transformations \bref{teta bos} and of the Wess-Zumino term
\bref{wz bos} indicates that
this system fits the requirements described in sect.4.
Indeed, since the commutator of a Weyl transformation and a diffeomorphism
$$
  [\delta_R(v^\alpha), \delta_W(\lambda)]=
  \delta_W(-v^\alpha\partial_\alpha\lambda),
$$
does not contain diffeomorphisms, the structure constants $T^A_{bB}$ vanish
and condition \bref{cond t} is trivially satisfied, while condition
\bref{first cond 2} is also seen to hold due to the Weyl invariance of the
kinetic operator $\Box$ in \bref{wz bos}.

\subsection{Extended proper solution, background term and
            extended regularization}

\hspace{\parindent}
Having verified that the application of the regularization process
described in sect.4 to this model is sensible, let us pass now to implement
it. First of all, from $S_{\rm ext}$ \bref{tilde s bos} and the Wess-Zumino
action \bref{wz bos}, we should recognize the relevant extended proper
solution $W_0$ and the background term $M_{1/2}$.
The canonical transformation \bref{teta trans} adapted to this case
performs this task and determines them to be
\bea
   W_0&=& \hat S(\Phi, K)
   +\int\dif^2\xi
   \left[\left(\frac{D-26}{48\pi}\right)
\left(\frac12\theta\Box\theta\right)+
  \theta^* \gh C^\alpha\partial_\alpha \theta\right],
\label{til w bos}\\
  M_{1/2}&=&-\int\dif^2\xi\,\left[\theta^*\gh C-
   \left(\frac{D-26}{48\pi}\right) \sqrt{g}\,R\,\theta\right],
\nonumber
\eea
which are seen to fulfill relations \bref{ncme}, \bref{inv back} and
\bref{n anom}.

The modified extended proper solution $\tilde W_0$, obtained from $W_0$
by leaving undetermined the numerical coefficients of the
kinetic operator for the extra variables,
is obtained through the substitution
$a=\left(\frac{D-26}{48\pi}\right)\rightarrow \widehat a$ in
\bref{til w bos}, i.e.,
\be
  \tilde W_0= \hat S(\Phi, K)
   +\int\dif^2\xi
   \left[\frac{\widehat a}{2}\,\theta\Box\theta+
  \theta^* \gh C^\alpha\partial_\alpha \theta\right].
\label{til w bos p}
\ee

The regulator $\gh R_\theta$ \bref{r teta}
for the $\theta$ sector is determined from
the new $\theta$ kinetic operator in \bref{til w bos p} once an explicit
mass matrix is chosen. In this case, the similarity of the kinetic
operators for the matter and $\theta$ sector suggests to use a similar
mass matrix for the PV field of the extra variable
$$
   T_{\theta}= T_{\theta,0}= \widehat a \sqrt g,
$$
from which, using expression \bref{r teta}, the following regulator
$\gh R_\theta$ is obtained
$$
   \gh R_\theta=\frac1{\sqrt g}\,\Box,
$$
i.e., exactly the same regulator as for the matter part \bref{reg corda},
with the only difference that now only one scalar field is involved.
Therefore,
the contribution of the extra sector to the antifield independent
part of the anomaly is just expression
\bref{weyl anom} with numerical coefficient $-1$. A first
effect of the extra degree of freedom at one loop is thus a
``renormalization''
of the original coefficient $a=\left(\frac{D-26}{48\pi}\right)$ of the
anomaly \bref{weyl anom} to $\left(\frac{D-25}{48\pi}\right)$ \cite{Fuj90}.

Now, cancelation of the antifield independent sector of
the $\hbar$ part in \bref{reg anomaly 2} is acquired for
$$
    \tilde M_{1/2}=-\int\dif^2\xi\,\left[\theta^*\gh C-
   \left(\frac{D-25}{48\pi}\right) \sqrt{g}\,R\,\theta\right],
$$
while the $\hbar$ equation $(\tilde M_{1/2}, \tilde W_0)=0$
is fulfilled by taking
$\widehat a=\left(\frac{D-25}{48\pi}\right)$ in \bref{til w bos p}. These
conditions together lead to the vanishing of the antifield
independent part of the
obstruction $\tilde{\gh A}$ \bref{reg anomaly 2} in the extended theory.

Finally, coming back to the original variables $\theta, \theta^*$,
these effects appear realized together as a shift of the original
coefficient of the Wess-Zumino term \bref{wz bos} in the same amount. The
final, one-loop renormalized Wess-Zumino action reads then
$$
   \tilde{M}^{(0)}_1(g_{\alpha\beta},\theta)=\int\dif^2\xi
   \left\{\left(\frac{D-25}{48\pi}\right)
\left[\frac12\theta\Box\theta+\sqrt{g}R\theta\right]\right\}.
$$
This result was
earlier obtained in \cite{DDK} and further reproduced in \cite{MM}, by
using a heat kernel regularization procedure for the non-trivial, gauge
invariant measure of the Liouville field, and in \cite{Fuj90} through the
application of the background field method. An alternative approach,
based in a canonical quantization of the regularized system taking into
account changes in the type of constraints, can also be found in
\cite{Andrei}.

\section{Conclusions and Outlook}

\hspace{\parindent}%
The aim of this paper has been to further study the formulation of the
extended Field-Antifield formalism for anomalous gauge theories,
previously proposed
and developed in \cite{GP93} \cite{Mars}, by means of the
incorporation of quantum one-loop effects coming from the extra
variables sector. To do that, an extension of the PV scheme proposed in
\cite{TNP89} \cite{T1} \cite{TP93} has been constructed to explicitly take
into account at once the regularization of both the original and the extra
fields, maintaining as far as possible the features characterizing the
(regularization of the) original theory. In this fashion, background terms,
known to appear in other formulations and giving rise to the anomalies in a
different way, as well as a new proper solution in the extended space,
directly arise from the combination $(\tilde S+\hbar \tilde M^{(0)}_1)$
of the extended, non proper solution and of the Wess-Zumino term
as a result of a canonical transformation in the extra field sector.
Unfortunately, only
a certain type of theories (bosonic string, abelian chiral Schwinger model,
etc.) seems to admit the perturbative description we present, indicating
that maybe a quantum treatment of Wess-Zumino terms goes beyond the scope
of the usual $\hbar$ perturbative expansion. In any case, our proposal
works for the restricted theories we study, yielding in the end
cancellation of the antifield independent part of the complete anomaly and
thus BRST invariance of the extended theory up to one-loop in this
restricted sector. Furthermore, in the particular example we
present --the bosonic string-- the application of this general framework
leads to a natural interpretation of the well-known shift
$(26-D)$ to $(25-D)$ as a one-loop renormalization of the
Wess-Zumino term due to the extra (Liouville) field quantum effects.
In any case, however, it should be stressed that the physical character of
an anomalous
gauge theory quantized in this fashion is not answered from the above
construction, although it would be of interest to analyze the unitarity of
the extended theory along the lines of
\cite{GP92}. There, it has been shown that, under certain conditions on the
generators of the gauge algebra, unitarity relies in the norm of
the classical gauge invariant degrees of freedom. In the present case
this would amount to study the norm of the "classical" degrees of
freedom associated with the new proper solution $W_0$.

The extended field-antifield formalism, on the other hand, presents some
other interesting features. Its covariance, for instance
--in the sense that the role of the anomalous propagating degrees of
freedom, played in previous approaches by the pure gauge
modes of the gauge fields (e.g., the conformal mode of the metric field in the
bosonic string), is now taken over by the extra degrees of freedom-- allows
to get rid of these pure gauge fields by gauge-fixing it as in the usual
non-anomalous theories. Related to this covariance, it would be of interest
to elucidate the relationship between our proposal and the description
given in \cite{TP93}, where the anomalous propagating degrees of freedom
are associated with some original antifields. Moreover, the extended
formalism also permits a complete determination of the transformation
properties of the extra variables and of the Wess-Zumino action from the
underlying (quasi)group structure, although the locality of such objects,
which relies on the locality of the quantitites defining the (quasi)group,
remains as an open problem worth to be investigated.

Finally, we would like to stress the importance that throughout our
developments the alternative expression for the anomaly \bref{a 2 quad}
involving only the regulator has found. Apart for its simplicity and the
fact that it stablishes a clear relationship between anomalies and
transformation properties of the regulator, the combination of its form
together with the expansion \bref{exp r} for a general regulator yields
the result that for ``closed'' theories
(i.e., $\delta^2_\Sigma=0$) anomalies obtained in the PV scheme
split into two $\delta$ off-shell invariant parts, one of them antifield
independent while the other one carries all the antifield dependence.
Algebraic counterexamples to this result has recently been given
in \cite{Brandt} in the form of solutions of the Wess-Zumino consistency
conditions with a non-trivial dependence on the antifields.
Our result excludes the appearence of such type of
anomalies and indicates that the regularization procedure
acts as a sort of ``selection rule'' identifying from the complete set of
algebraic solutions of the consistency conditions a subset of
``physically'' realized anomalies. Further understanding of this point
clearly deserves future investigation.

\section*{Acknowledgements}

\hspace{\parindent}%
We would like to acknowledge A. van Proeyen, W. Troost, F. de Jonghe,
R. Siebelink and S. Vandoren for many fruitful discussions and also
A. van Proeyen, W. Troost and F. de Jonghe for the
critical reading of the manuscript. J.\,G. is grateful to Prof. S. Weinberg
for the warm hospitality at the Theory Group of the University of Texas at
Austin, where part of this work was done.

This work has been partially supported by
the Robert A. Welch Foundation, NST Grant 9009850,
CICYT project no.\,AEN93-0695 and
NATO Collaborative Research Grant 0763/87.

\appendix

\section{Alternative form of $(\Delta \hat S)_{\rm reg}$}

\hspace{\parindent}%
Our main purpose here is to show that expression \bref{a 2 quad}
for $(\Delta \hat S)_{\rm reg}$ is completely equivalent to the form
obtained in \cite{T1} \cite{TP93}
\be
   (\Delta \hat S)_{\rm reg}=
   \left[\left(K^A_{\,B}+\frac12 (T^{-1})^{AC}\delta T_{CB}(-1)^B
   \right)
   \left(\frac{1}{1-R/M}\right)^B_{\,A}\right],
\label{delta s reg}
\ee
with $K^A_{\;B}$ defined from \bref{s der} to be
$$
  K^A_{\;B}=\left(\lder{}{K_A}\rder{}{\Phi^B}\hat{S}(\Phi, K)\right).
$$

The idea consists in using some relations between
$K^A_{\;B}$, $T_{AB}$ and $R^A_{\;B}$ and express \bref{delta s reg}
in terms of the regulator $R^A_{\;B}$ alone.
First, the symmetry properties of
$T_{AB}$ and the definition of the supertranspose of
$K^A_{\;B}$
$$
   T_{BA}=(-1)^{A+B+AB} T_{AB},\quad\quad
   (K^T)^{\;A}_B=(-1)^{A(B+1)}K^A_{\;B},
$$
allows to rewrite \bref{delta s reg} in the equivalent form
\be
   (\Delta \hat S)_{\rm reg}=
  \left\{\frac12
  \left[\left(K+ R^{-1}T^{-1}K^T TR\right)^A_{\;B} +
   (T^{-1})^{AC}\delta T_{CB}(-1)^B\right]
   \left(\frac{1}{1-R/M}\right)^B_{\;A}\right\}.
\label{new a quad}
\ee

Now, differentiation of the classical master equation for $\hat S$
$$
  \lder{}{\Phi^A}\rder{}{\Phi^B}(\hat S,\hat S)=0,
$$
provides the following identity between $K^A_{\;B}$ and $(TR)_{AB}$
$$
  (TRK)_{AB}+(TRK)_{BA}(-1)^{AB}+(-1)^B\delta (TR)_{AB}=0.
$$
This relation can now be used to express the combination
$(K+R^{-1}T^{-1}K^T TR)^A_{\;B}$ in \bref{new a quad} as
$$
   (K+R^{-1}T^{-1}K^T TR)^A_{\;B}=
    -(R^{-1})^A_{\;C}\delta R^C_{\;B}(-1)^B
   -(R^{-1}T^{-1})^{AC}\delta T_{CD} (-1)^D R^D_{\;B},
$$
which substituted in \bref{new a quad} eliminates its explicit dependence
on $K^A_{\;B}$ and $T_{AB}$. In the end, use of
the definition of the supertrace
as well as its cyclic property in the resulting expression yields
\bref{a 2 quad}, as we would like to show.

\section{Counterterms}

\hspace{\parindent}%
The particular expression of the anomaly is not unique. Its form depends
both on the
intermediate regularization scheme (i.e., on the mass part of
$S_{\rm PV}$ \bref{orig pv action}, for example) and on the form of the
counterterm $M_1$.
Different expressions of the consistent anomaly, or equivalently, of
$(\Delta \hat S)_{\rm reg}$, are thus expected to be related by the BRST
variation of a local counterterm. The form of this counterterm
when the regularization schemes are connected
by a continuous interpolating regulator was conjectured in \cite{TNP89}.
Here, we derive the expression of this counterterm%
\footnote{An alternative proof of this conjecture can be found in
\cite{Frank}.}.

Consider two different regularizations defined in terms of the mass
matrices and regulators $(T_0, R_0)$ and $(T_1, R_1)$,
satisfying $T_0R_0=T_1R_1=(TR)$.
In each one, \bref{a 2 quad} reads
\be
   (\Delta \hat S)_{\rm reg,0(1)}=
   \delta\left\{-\frac12{\rm Tr}
   \ln\left[\frac{(TR)}{T_{0(1)}M-(TR)}\right]\right\}.
\label{un dos}
\ee
Assume now that there exists a continuous path $T(t)$, $t\in[0,1]$,
interpolating from the first mass matrix, $T_0=T(0)$, to the second,
$T_1=T(1)$. This interpolation induces in turn another interpolation
between $R_0$ and $R_1$
\be
   T(t)R(t)=(TR)\Rightarrow
   R(t)=T^{-1}(t)(TR).
\label{rdet}
\ee

The difference between the two regularized expressions of $\Delta \hat S$
\bref{un dos} is then
$$
   (\Delta \hat S)_{\rm reg,1}-(\Delta \hat S)_{\rm reg,0}=
  \delta\left\{-\frac12{\rm Tr}
   \ln\left[\frac{(TR)}{T_{1}M-(TR)}\right]
   +\frac12{\rm Tr}
   \ln\left[\frac{(TR)}{T_{0}M-(TR)}\right]\right\}=
   \delta M_0(\Phi,K),
$$
from which we can read off the form of the counterterm
relating them. Further use of $T(t)$
allows to rewrite $M_0$ as an integral over $t$
\be
    M_0(\Phi,K)=
    \int^1_0\dif t\, \partial_t\left\{-\frac12{\rm Tr}
   \ln\left[\frac{(TR)}{T(t) M-(TR)}\right]\right\}
    = \int^1_0\dif t\, {\rm Tr}\left[\frac12(T^{-1}(t)\partial_t T(t))
    \frac{1}{(1-R(t)/M)}\right],
\label{mzero t}
\ee
which exactly coincides
with the expression conjectured in \cite{TNP89}.

In view of the results of sec.5, it is convenient to rewrite
\bref{mzero t} as an expression involving only $R(t)$.
This can be achieved by considering
the independence of the kinetic term $(TR)$ for the
PV fields on $t$ \bref{rdet}. This property induces the relation
$$
   \partial_t[T(t)R(t)]=\partial_t(TR)=0\Leftrightarrow
   (T^{-1}(t)\partial_t T(t))=-(\partial_t R(t))R^{-1}(t),
$$
which, when substituted in \bref{mzero t}, yields the desired
expression
\be
   M_0(\Phi,K)=\int^1_0\dif t\, {\rm Tr}
   \left[-\frac12 (R^{-1}(t)\partial_t R(t))
   \frac{1}{\left(1-R(t)/M\right)}\right].
\label{m zero r}
\ee

\section{Transformation properties of the regulator $\gh R_\theta$}

\hspace{\parindent}%
In sect.4.4 we argued that the introduction of extra variables could lead
to type $A$ anomalies unless an invariant regulator
$\gh R_\theta(\phi)$ is used.
The purpose of this appendix is the obtention of
the conditions ensuring the vanishing of these extra anomalies.

The transformation properties of the regulator
$ (\gh R_\theta)^a_{\;b}(\phi)=
   -i(T^{-1})^{ac} \widehat{D}_{cb}(\phi),
$
\bref{r teta}
can be obtained from those of the kinetic term
$\widehat{D}_{cb}(\phi)$ \bref{d wide}
and the mass matrix
$(T^{-1}_{0})^{ab}\equiv(T^{-1})^{ab}$.
In particular, since
$\widehat{D}_{cb}(\phi)$ \bref{d wide}
only differs from the original
one in its numerical coefficients, the transformation
properties of the former are completely determined from that
of the latter \bref{first cond 2}, \bref{dab trans}. We have thus
\be
   \delta_{(a)}\widehat D_{ab}=0,\quad\quad
   \delta_{(A)} \widehat{D}_{ab}=
   \left(\widehat{D}_{ac} T^c_{bB}+
         \widehat{D}_{bc} T^c_{aB}\right) \veps^B.
\label{trans wide}
\ee
It is obvious then that the vanishing
of type $A$ anomalies
lies entirely on the $A$ transformation of $T_{ab}$.
Indeed, taking into account
\bref{trans wide}, the $a$ transformation of the
regulator reads
\be
   \delta_{(a)} (\gh R_\theta)^a_{\;b}=
   -i[\delta_{(a)}(T^{-1})^{ac}] \widehat{D}_{cb}.
\label{delta a r}
\ee
In this way, new contributions to the
original $A_a(\phi)$ anomalies only come from the noninvariance of $T_{ab}$
under the $a$ part.
On the other hand, the $A$ transformations for $\gh R_\theta$ are
\bea
   \delta_{(A)} (\gh R_\theta)^a_{\;b}&=&
   -i[\delta_{(A)}(T^{-1})^{ac}] \widehat{D}_{cb}
   -i(T^{-1})^{ac} [\delta_{(A)}\widehat{D}_{cb}]
\nonumber\\
  &=& i (T^{-1})^{ae} (\delta_{(A)}T_{ed})(T^{-1})^{dc}
 \widehat{D}_{cb} -i(T^{-1})^{ac}
   \left(\widehat{D}_{cd} T^d_{bB}+
         \widehat{D}_{bd} T^d_{cB}\right) \veps^B.
\nonumber
\eea

Hence, to avoid new anomalies, the above
transformation should be zero or, at least,
of the form $\left[\gh R_\theta, G\right]$.
This is precisely the case if
\be
   \delta_{(A)} T_{ab}=
   \left(T_{ac} T^c_{bA}+ T_{bc} T^c_{aA}\right)\veps^A,
\label{trans tab}
\ee
yielding in this way the transformation rule
$$
   \delta_{(A)} (\gh R_\theta)^a_{\;b}=
    [\gh R_\theta, G]^a_{\;b},\quad\quad\mbox{with}\quad\quad
    G^a_{\;b}\equiv T^a_{Bb}\veps^B.
$$
In summary, \bref{trans tab} is the suitable transformation property of
$T_{ab}$ considered in sect.4.4 guaranteeing that no new symmetries
become anomalous upon introduction of the extra variables.

With respect to the completely anomalous case (i.e., the abelian theory)
the transformation of the regulator is as in \bref{delta a r} with
the substitution $a\rightarrow\alpha$, i.e.,
$$
   \delta (\gh R_\theta)^\alpha_{\;\beta}=
   -i[\delta(T^{-1})^{\alpha\gamma}] \widehat{D}_{\gamma\beta},
$$
so that an extra contribution to the original anomaly is likely to appear
if $T_{\alpha\beta}$ is not invariant. In the
usual cases, a constant mass matrix without any dependence on the fields
can be chosen. Then,
$\delta(T^{-1})^{\alpha\beta}=0$, and no new contributions
arises. In any case, the existence of such invariant and/or constant
mass matrix should be analyzed model by model.

\endsecteqno

\end{document}